\documentclass[11pt]{article}
\usepackage{geometry}
\usepackage{authblk}
\usepackage{xcolor,amssymb}
\usepackage{graphicx}
\usepackage{booktabs}
\usepackage[sorting=none]{biblatex}
\newcommand\pubnumber{SLAC-PUB-17659}
\newcommand\pubdate{\today}
\newcommand\pubblock{\rightline{\begin{tabular}{l} \pubnumber\\
          \pubdate \end{tabular}}}

\title{\textbf{XCC: An X-ray FEL-based $\gamma\gamma$ Collider Higgs Factory}}

\author[1,a]{Tim Barklow}
\author[1]{Su Dong}
\author[1]{Claudio Emma}
\author[1]{Joseph Duris}
\author[1]{Zhirong Huang}
\author[1]{Adham Naji}
\author[1]{Emilio Nanni}

\author[2]{James Rosenzweig}
\author[1]{Anne Sakdinawat}
\author[1]{Sami Tantawi}
\author[1]{Glen White}
\affil[1]{SLAC Linear Accelerator Center, Stanford, Menlo Park, CA}
\affil[2]{Particle Beam Physics Laboratory, University of California Los Angeles, CA}
\affil[a]{timb@slac.stanford.edu}
\date{}

\begin{document}

\begin{flushright}
\pubblock
\end{flushright}

\begingroup
\let\newpage\relax% Void the actions of \newpage
\maketitle
\endgroup

\begin{center}
\section*{Abstract}
\end{center}
{

   This report describes the design of a $\gamma\gamma$ Higgs factory in which 
  $62.8$~GeV electron beams collide with  1~keV X-ray free electron laser (XFEL) beams to produce colliding beams of 62.5~GeV photons. The Higgs boson production rate is 
  34,000 Higgs bosons per $10^7$ second year, roughly the same as the ILC Higgs rate. The electron accelerator is based on cold copper distributed coupling (C$^3$) accelerator  technology.  The 0.7~J pulse energy of the XFEL represents  a 300-fold increase over the pulse energy of current soft x-ray FEL's.  Design challenges are discussed, along with the R\&D to address them, including demonstrators.
  %In contrast, the distribution for an $x=4.82$ collider would have a peak at the Higgs boson mass with widths of 22~GeV (5.6~GeV) on the low (high) side plus additional structure at lower $\gamma\gamma$ center-of-mass energies, and would therefore produce a much greater  $\gamma\gamma$ background to the Higgs signal.
  %With such a unique 
  %experimental environment the Higgs physics output of an XFEL $\gamma\gamma$ Higgs factory is
  %greater than that of  optical wavelength $\gamma\gamma$ colliders.
  }
%\vspace{7.0cm}
\vspace*{\fill}

\begin{center}
\textit{Contributed to}  \textit{The US Community Study
on the Future of Particle Physics (Snowmass 2021)} \\
\textit{Seattle, WA} \\
\textit{July 17 - 26, 2022} \\
\end{center}

%\vspace{0.5cm}\rule{0.5\textwidth}{0.4pt}\vspace{0.5cm}

%\end{center}

\newpage

\tableofcontents

\newpage

\section{Introduction}

\subsection{Concept Overview}
To date, $\gamma\gamma$ collider Higgs factory designs have utilized optical wavelength lasers\cite{Ginzburg:1981vm}\cite{Ginzburg:1982yr}\cite{Telnov:1989sd}\cite{Asner:2001ia}\cite{Asner:2001vh}.  The
center-of-mass energy of the electron--photon system is usually constrained to $x<4.82$, where $x=4E_{e}\omega_0/m^2_e$, $m_e$ is the electron mass and $E_{e}$ ($\omega_0$) is the electron (laser photon) energy.  Larger $x$ values are problematic due to the linear QED thresholds of $x=4.82$  ($x=8.0$) for the processes $\gamma\gamma_0\rightarrow e^+e^-$  ($e^-\gamma_0\rightarrow e^-e^+e^-$), where $\gamma$ and $\gamma_0$ refer to the Compton-scattered and laser photon, respectively.  Larger $x$ values, however, also carry advantages.  As $x$ is increased,
the $\gamma\gamma$  luminosity distribution with respect to center-of-mass energy is more sharply peaked near the maximum center-of-mass energy value. Such a distribution increases the production rate of a narrow resonance relative to $\gamma\gamma$ background processes when the peak 
is tuned to the resonance mass.

\begin{figure}
\centering
     \includegraphics[width=1.00\textwidth]{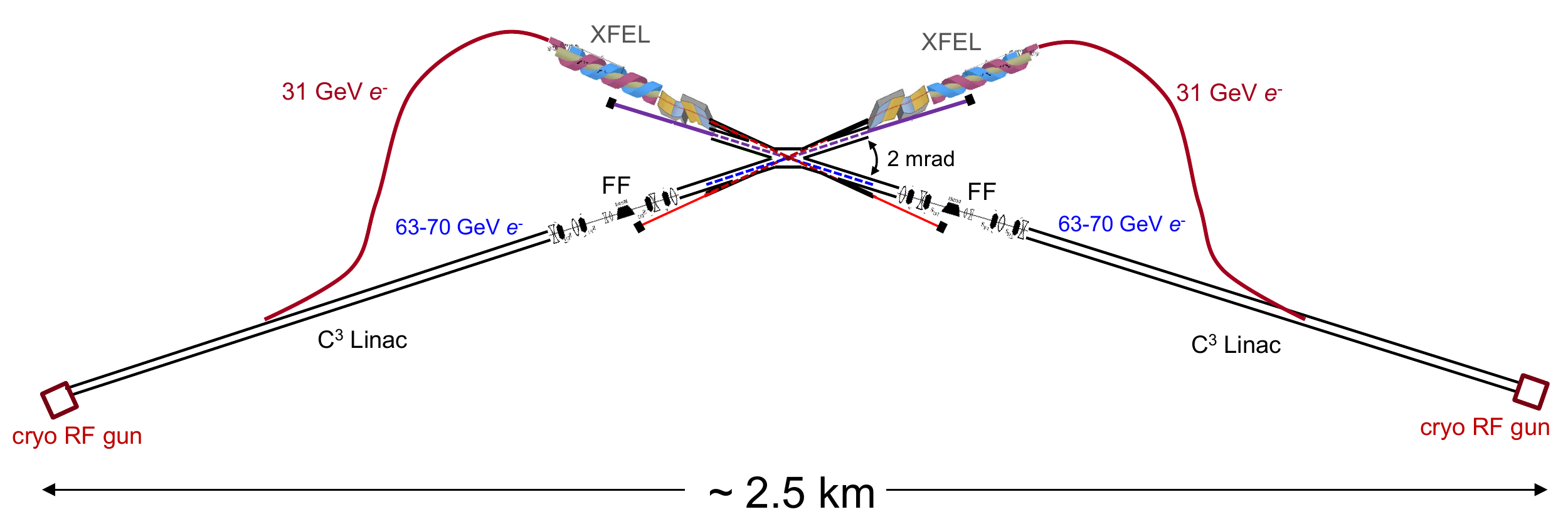}
     \caption{Schematic of XCC including cryogenic RF injector, C$^3$ Linac, electron beam final focus (FF) and XFEL.}
     \label{fig:schematic}
 \end{figure}

A schematic of the $\gamma\gamma$ collider, or XFEL Compton Collider (XCC),
is shown in Fig.~\ref{fig:schematic}.   A low emittance cryogenic RF gun produces 90\% polarized electrons with $0.62\times 10^{10}$ electrons per bunch and 76 bunches per train at a repetition
rate of 240~Hz.   The normalized horizontal and vertical gun emittances are 0.12 microns each.  A linear accelerator (Linac) utilizing  cold copper distributed coupling (C$^3$) technology\cite{Bane:2018fzj} accelerates the electron bunches with a bunch spacing of 5 ns and a gradient of 70 MeV/m.  At the 31~GeV point every other bunch is diverted to the XFEL line where a helical undulator produces circular polarized 1~keV X-ray light with 0.7 Joules per pulse.   The remaining bunches continue down the Linac until reaching an energy of 62.8~GeV, after which they pass through a final focus section that squeezes the geometric horizontal and vertical spot sizes to 5.4~nm at the primary
$e^-e^-$ interaction point (IP).  The $e^-e^-$ geometric luminosity is $9.7\times 10^{34} \rm{cm}^2\ \rm{s}^{-1}$.  
At the Compton interation point (IPC), the 62.8 GeV electrons collide with the X-ray laser light from the opposing XFEL line to produce 62.5~GeV photons.  The X-ray light has been focused at this point from $9\ \mu$m at the end of the XFEL to a waist radius of $a_\gamma$=30~nm using a series of Kirkpatrick-Baez (KB) mirrors.

\begin{figure}
\centering
     \includegraphics[width=0.75\textwidth]{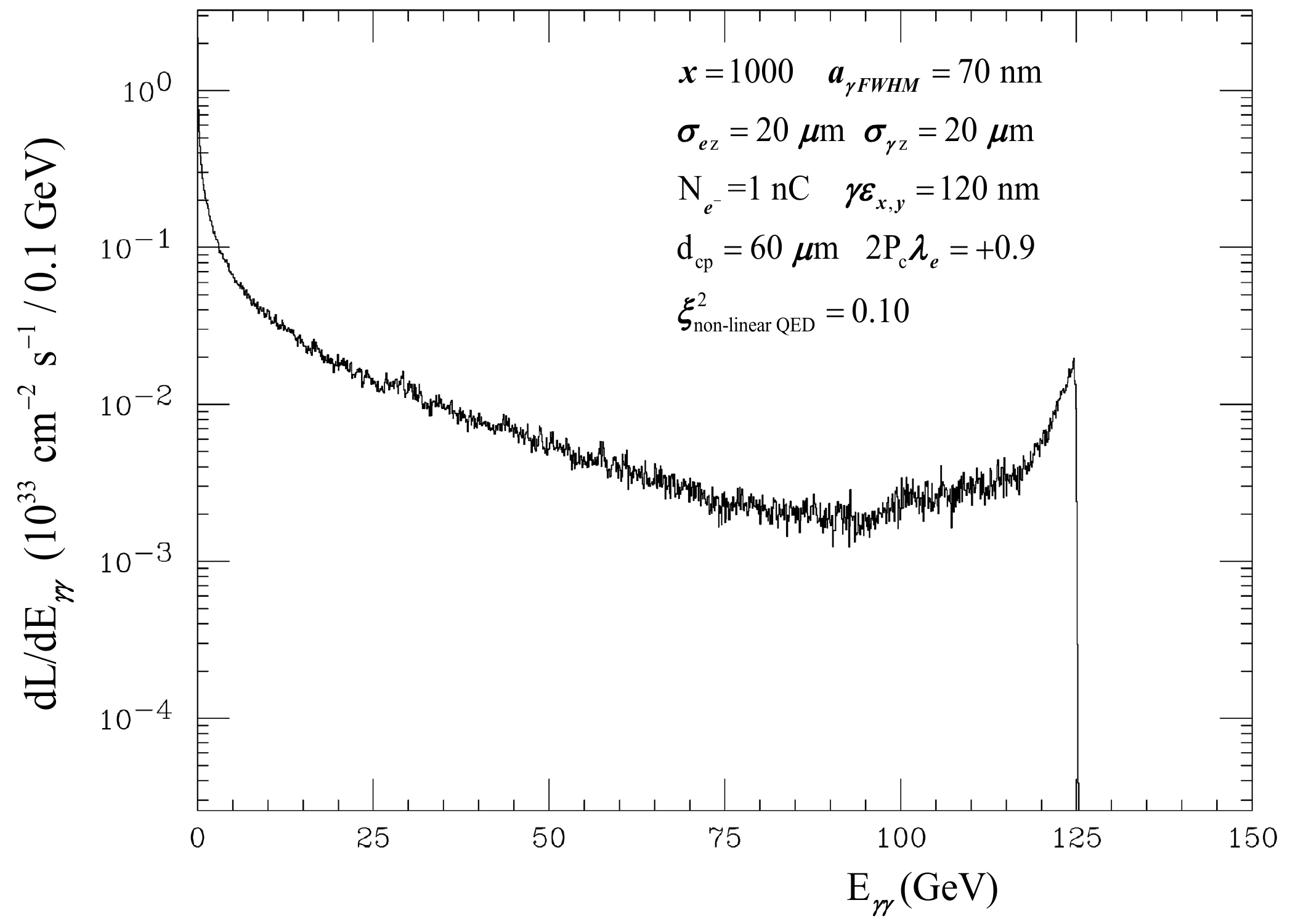}
     \includegraphics[width=0.75\textwidth]{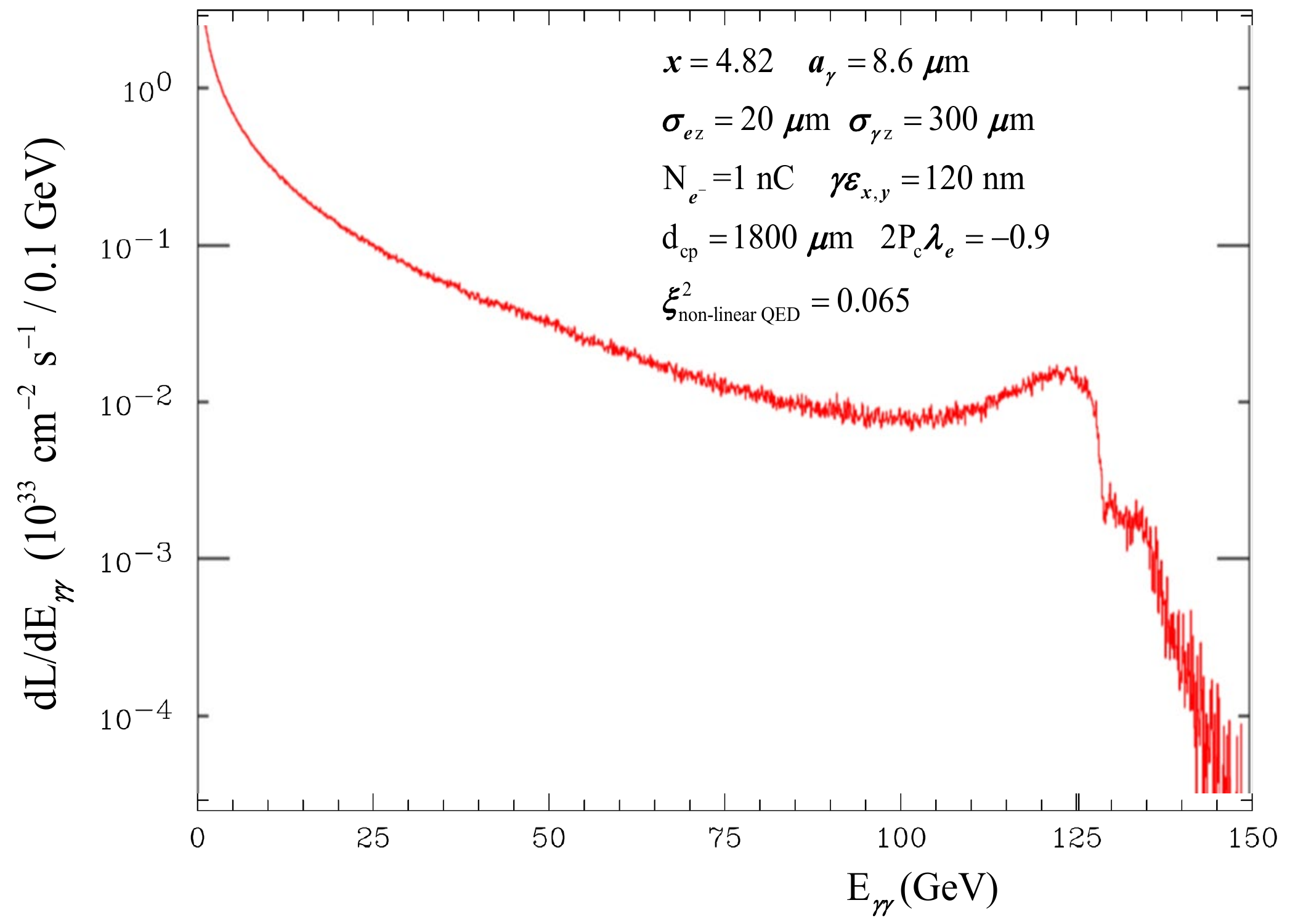}
     \caption{$\gamma\gamma$ luminosity as a function of $\gamma\gamma$ center-of-mass energy $E_{\gamma\gamma}$ for x=1000 \& $2P_c\lambda_e=+0.9$ (top) versus x=4.82 \& $2P_c\lambda_e=-0.9$ (bottom) as calculated by the  CAIN MC, where $P_c$ and $\lambda_e$ are the helicities of the laser photon and electron, respectively. The x-ray waist radius $a_\gamma$ at the Compton interaction point and the electron (x-ray) beam r.m.s longitudinal sizes, $\sigma_{ez}$ ($\sigma_{\gamma z}$), are indicated.  The 0.05\% electron beam energy spread, linear QED Bethe-Heitler scattering  ($e^-\gamma_0\rightarrow e^-e^+e^-$) and non-linear QED effects in Compton scattering ($e^-\gamma_0\rightarrow e^-\gamma$) and Breit-Wheeler scattering ($\gamma\gamma_0\rightarrow e^+e^-$) are included in the CAIN simulation, where $\gamma$ and $\gamma_0$ refer to the Compton-scattered and laser photon, respectively.
     }
     \label{fig:x1000lumiplus}
 \end{figure}

The distribution of $\gamma\gamma$ luminosity versus $\gamma\gamma$ center-of-mass energy $E_{\gamma\gamma}$ as calculated with the CAIN Monte Carlo\cite{Chen:1994jt} is shown in Fig.~\ref{fig:x1000lumiplus}
for $2P_c\lambda_e=+0.9$, where $P_c=+1$ and $\lambda_e=+0.45$ are the helicities of the  laser photon and electron, respectively. For comparison, the corresponding distribution from an x=4.82 optical laser $\gamma\gamma$ collider (OCC) is also shown.  The OCC --  presented here solely as an optical laser $\gamma\gamma$ collider counter-example to the XCC -- has the same parameters as XCC except that the XFEL is replaced with the optical laser in~\cite{Telnov_2020}, the electron beam energy is increased from 62.8~GeV to 86.5~GeV,  the distance $d_{cp}$ between the IPC and IP has been increased from $60\ \mu$m to $1800\ \mu$m, and $2P_c\lambda_e=-0.9$.  The  distribution for x=1000 has an asymmetric peak at the Higgs boson mass with a leading edge width of 0.3~GeV.  In contrast, the $x=4.82$ distribution has a peak at the Higgs boson mass with a leading edge width of 3.5~GeV and a long high-side tail due to multi-photon non-linear QED Compton scattering, characterized by the parameter $\xi^2=2n_\gamma r_e^2\lambda/\alpha$  where $n_\gamma$ is the laser photon density, $r_e$ is the classical electron radius, and $\lambda$ is the laser photon wavelength.  Although the non-linear QED parameter $\xi^2=0.10$ for the $x=1000$ configuration is 50\% larger than that of the $x=4.82$ configuration, the long high-side non-linear QED tail is absent because the 
difference between the maximum linear and non-linear QED photon energies, $E_e/(x+1)$, is very small.   The large low-side tail in the luminosity distributions for $0<E_{\gamma\gamma}<100$~GeV is due to beamstrahlung radiation.

%Relative to the optical laser x=4.82 collider, the XFEL x=1000 collider provides improved Higgs signal-to-background and a much better total Higgs width measurement 
%(30 MeV vs. 500 MeV for a dedicated one year energy scan around the  Higgs mass).
%, and would therefore produce a much greater  $\gamma\gamma$ background to the Higgs signal.

The total luminosity and the luminosity for $\sqrt{\hat{s}}>100$~GeV for $\gamma\gamma$
and other processes are listed in Table~\ref{tab:lumisummary} for the XCC running in $\gamma\gamma$ mode at $\sqrt{s}=125$~GeV. The $\gamma \gamma$ luminosity of $1.2\times 10^{33}$ cm$^{-2}$ s$^{-1}$ for $\sqrt{\hat{s}}>100$~GeV may seem low compared to other proposed Higgs factories.  However, due to the narrow Higgs resonance the Higgs production rate at XCC is that of a $10^{34}$ cm$^{-2}$ s$^{-1}$ $e^+e^-$ collider, as shown in Table~\ref{tab:higgsratesummary}.  In lieu of a background estimate from Monte Carlo event generation and detector simulation, the number of hadronic events with $\sqrt{\hat{s}/s}>0.4$ is chosen as a measure of the background rate. The background rate for the XCC is comparable to that of the ILC, and is dominated by beamstrahlung luminosity. 
%In point of fact, without the beamstrahlung luminosity the XCC background would be significantly less than that of the ILC.     
From Table~\ref{tab:higgsratesummary} and  Fig.~\ref{fig:x1000lumiplus} it can be seen that the XCC background rate is significantly smaller than that of the OCC optical $\gamma \gamma $ collider.

 \begin{table}

    \centering
    \begin{tabular}{    c  | c | c  } \toprule
        \multicolumn{3}{ c}{$\gamma\gamma$ mode $\sqrt{s}=125$~GeV} \\ \midrule
       & \multicolumn{2}{ c}{Luminosity ($10^{34}$ cm$^{-2}$ s$^{-1}$)} \\[3pt]
        Process  & Total &  $\sqrt{\hat{s}}>100$~GeV \\ \midrule
     $\gamma\gamma$  & 2.1 & 0.12 \\
     $e^-e^-$        & 0.23 & 0.18 \\
     $e^-\gamma+\gamma e^-$    & 2.5 & 0.42 \\
     $e^+e^-+e^-e^+$        & 0.48 & 0.05 \\
     $e^+\gamma+\gamma e^+$     & 0.47 & 0.01 \\
        \bottomrule
    \end{tabular}
        \caption{ \label{tab:lumisummary}Total luminosity and luminosity for $\sqrt{\hat{s}}>100$~GeV for different processes at the XCC running in $\gamma\gamma$ mode at $\sqrt{s}=125$~GeV.}
\end{table}

 \begin{table}

    \centering
    \begin{tabular}{    c  | c c c c c c} \toprule
     Machine   &  $E_{e^-}$ (GeV) & $N_{e^-}$ (nC) & Polarization & $N_{\rm H}$/yr 
     & $N_{\rm Hadronic}/N_{\rm H}$     
%     & $N_{\rm Had}$ $(\sqrt{\hat{s}/s}>0.4)/N_{\rm H}$ %

     &   $N_{\rm minbias/BX}$ \\ \midrule
      XCC & 62.8 & 1.0 & 90\% $e^-$  & 34,000  & 170  & 9.5 \\ \midrule
      OCC & 86.5 & 1.0 & 90\% $e^-$ & 30,000 & 540 & 50  \\ \midrule
      ILC & 125 & 3.2 & -80\% $e^-$  +30\% $e^+$ & 42,000  & 140  & 1.3 \\
      ILC & 125 & 3.2 & +80\% $e^-$  -30\% $e^+$ & 28,000  & 60  & 1.3 \\

        \bottomrule
    \end{tabular}
        \caption{ \label{tab:higgsratesummary} Higgs rate $N_{\rm H}$/yr and background rate $N_{\rm Hadronic}/N_{\rm H}$ for the XCC, the OCC alternative $\gamma\gamma$ collider example, and the ILC.  A year is defined as $10^7$~s, and $N_{\rm Hadronic}$ refers to the number of hadronic events with $\sqrt{\hat{s}/s}>0.4$. The number of minimum bias events per bunch crossing is also indicated.}
\end{table}

\subsection{Higgs Physics}\label{sec:higgsphysics}
The XCC will measure $\sigma(\gamma\gamma\rightarrow H)\times \rm{BR}(H\rightarrow X)\ \propto \Gamma_{\gamma\gamma}\Gamma_{X}/\Gamma_{\rm tot}$ in  $H$ resonance production at $\sqrt{s_{\gamma\gamma}}=125$~GeV for a variety of Higgs decay modes $X= b\bar{b},\ c\bar{c},\ WW^*$, etc.  Given the entries in Table~\ref{tab:higgsratesummary}, the XCC errors for $\sigma(\gamma\gamma\rightarrow H)\times \rm{BR}(H\rightarrow X)$ should be  comparable to those of ILC.

\begin{figure}
\centering
     \includegraphics[width=0.75\textwidth]{Figures/x1000LumiSpectrum2pcle+0.9.pdf}
     \includegraphics[width=0.75\textwidth]{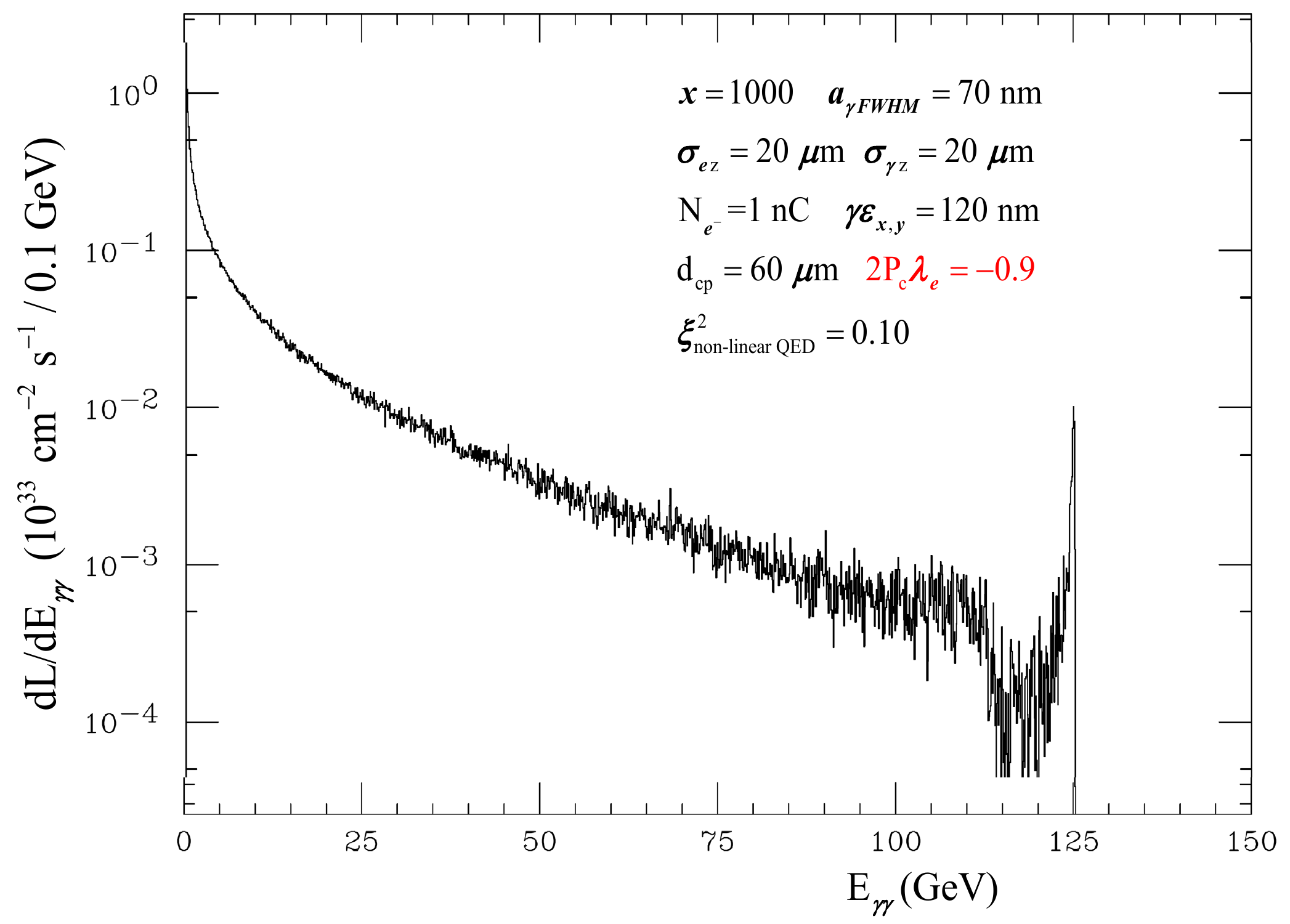}
     \caption{$\gamma\gamma$ luminosity as a function of $\gamma\gamma$ center-of-mass energy $E_{\gamma\gamma}$ for x=1000 \& $2P_c\lambda_e=+0.9$ (top) versus $2P_c\lambda_e=-0.9$ (bottom) as calculated by the  CAIN MC, where $P_c$ and $\lambda_e$ are the helicities of the laser photon and electron, respectively.  The x-ray waist radius $a_\gamma$ at the Compton interaction point and the electron (x-ray) beam r.m.s longitudinal sizes, $\sigma_{ez}$ ($\sigma_{\gamma z}$), are indicated. The 0.05\% electron beam energy spread, linear QED Bethe-Heitler scattering and non-linear QED effects in Compton and Breit-Wheeler scattering are included in the CAIN simulation.
     }
     \label{fig:x1000lumiminus}
 \end{figure}

To measure individual partial widths, an independent measurement of either $\Gamma_{\gamma\gamma}$
or $\Gamma_{\rm tot}$ is required. 
The total Higgs width $\Gamma_{\rm tot}$ can be measured with a scan of the XCC energy across the Higgs resonance. An alternate XCC polarization $2P_c\lambda_e=-0.9$ provides
a narrower leading edge width of 0.045~GeV (dominated by the 0.05\% $e^-$ beam energy spread) as shown in as shown in Fig.~\ref{fig:x1000lumiminus}. This comes at a cost, however, as the Higgs rate is reduced to 20,000 Higgs per year.  With such a configuration an energy scan can be used to measure the total Higgs width to an accuracy of $\Delta \Gamma_{\rm tot}=4.5$~MeV.  This is fine if $\Gamma_{\rm tot}$ is a few 10's of MeV, but clearly insufficient in the more likely event that the total Higgs width is close to its Standard Model value $\Gamma_{\rm tot}\approx4.0$~MeV.

The $\gamma\gamma$ partial width $\Gamma_{\gamma\gamma}$ can be measured directly in the process $e^-\gamma\rightarrow e^- H$ at $\sqrt{s}=140$~GeV.  The signal is a monochromatic 14.2~GeV electron, predominantly in the forward direction.  In order to achieve model independent ILC-like precision for Higgs couplings
and the Higgs total width, about one  $e^-\gamma\rightarrow e^- H$ event must be detected at $\sqrt{s}=140$~GeV for every 125 $\gamma\gamma\rightarrow H$ events collected at $\sqrt{s}=125$~GeV.  The cross section for Higgs production in $e^-\gamma$ collisions at $\sqrt{s}=140$~GeV is $\sigma(e^-\gamma\rightarrow e^- H)=4.1$~fb, assuming forward detector coverage down to an angle of 3~mrad.  With the current  $e^-\gamma$ design,
the yearly $e^-\gamma$ luminosity with $\sqrt{\hat{s}}$ within 1\% of the  140~GeV peak is 32.2~fb$^{-1}$.  Thus,  for every year collecting Higgs events at $\sqrt{s}=125$~GeV,
two years must be spent detecting $\sigma(e^-\gamma\rightarrow e^- H)$ at $\sqrt{s}=140$~GeV.

With a luminosity upgrade that doubles the Higgs production rates shown in Table~\ref{tab:higgsratesummary}, the ILC produces a total of $0.5\times 10^{6}$ $e^+e^-\rightarrow ZH$ events at $\sqrt{s}=250$~GeV over a $\sim10$~year time period.  The XCC could also produce   $0.5\times 10^{6}$ $\gamma\gamma \rightarrow H$  events over a 10~year period with a doubling of its luminosity.  However, the XCC must also produce 4000 $\sigma(e^-\gamma\rightarrow e^- H)$ events to go along with the $0.5\times 10^{6}$ $\gamma\gamma \rightarrow H$  events.   To achieve the required event samples in a reasonable amount of time, it is assumed that the XCC will be able to upgrade its luminosity by a factor of 3.8 by, for example, increasing the number of colliding bunches per train from 38 to 145.  With that scenario, the XCC can achieve the Higgs coupling precision shown in Table~\ref{tab:higgscouplings} in a 12~year period (4~years running $\gamma\gamma \rightarrow H$ at $\sqrt{s}=125$~GeV and 8~years with $e^-\gamma\rightarrow e^- H$ at $\sqrt{s}=140$~GeV)\cite{xccphysics}.

 \begin{table}

    \centering
    \begin{tabular}{    l  | c c } \toprule
    & ILC & XCC \\
     coupling $a$   &  $\Delta a$ (\%) &   $\Delta a$ (\%)  \\ \midrule
      $HZZ$ & 0.57 & 1.2 \\
      $HWW$ & 0.55 & 1.2 \\ 
      $Hbb$ &  1.0 & 1.4 \\ 
      $H\tau\tau$ & 1.2 & 1.4 \\ 
      $Hgg$ & 1.6 & 1.7 \\ 
      $Hcc$ & 1.8 & 1.8 \\ 
      $H\gamma\gamma$ & 1.1 & 0.77 \\ 
      $H\gamma Z$ & 9.1 & 10.0 \\ 
      $H\mu\mu$ & 4.0 & 3.8 \\ \midrule
      $\Gamma_{\rm tot}$ & 2.4 & 3.8 \\
      ${\Gamma_{\rm inv}}^\dagger$ & 0.36 & -- \\ 
      ${\Gamma_{\rm other}}^\dagger$ & 1.6 & 2.7 \\

        \bottomrule
    \end{tabular}
    
    $^\dagger$~95\% C.L. limit \qquad \qquad \qquad \qquad

        \caption{ \label{tab:higgscouplings} Higgs coupling precision for the full programs at ILC and XCC, as calculated with an Effective Field Theory (EFT) Higgs coupling fitting program\cite{Barklow:2017suo,Barklow:2017awn}. The full program at ILC includes $0.5\times 10^{6}$ $e^+e^-\rightarrow ZH$ events at $\sqrt{s}=250$~GeV.  For XCC there are  $0.5\times 10^{6}$ $\gamma\gamma \rightarrow H$  events at $\sqrt{s}=125$~GeV  and 4000 $e^-\gamma\rightarrow e^- H$ at $\sqrt{s}=140$~GeV. The errors on 
$\sigma(\gamma\gamma\rightarrow H)\times \rm{BR}(H\rightarrow X)$ are taken from~\cite{Bambade:2019fyw}, and are assumed to be the same for ILC and XCC for all decay modes $H\rightarrow X$ except $H\rightarrow {\rm invisible}$.}
 \end{table}

\subsection{Cost: XCC vs. $e^+e^-$ Higgs Factories}

 Compared to $e^+e^-$ Higgs factory proposals such as the ILC and  C$^3$-250\cite{Bai:2021rdg}, the XCC adds two additional beamlines and collision points, requires significant extensions of XFEL technology, produces Higgs physics that is comparable to -- but still not quite as good as --  the $e^+e^-$ Higgs factories, and must be run in both  $\gamma\gamma$ and $e^-\gamma$ initial state configurations.  Furthermore, a high energy physics $\gamma\gamma$ collider has never been built.
Nevertheless, the XCC concept is worth pursuing because every $e^+e^-$ linear collider proposal to date has been rejected due to its high cost.  Given that the XCC has no damping rings and that its beam energy is half that of $e^+e^-$ Higgs factories, the XCC may provide significant cost savings, perhaps enough to provide the difference between
Higgs factory approval and rejection.

Using the C$^3$-250 cost model\cite{Bai:2021rdg}, an initial estimate of the XCC cost breakout is given in Table~\ref{tab:capitalcost}.   When compared to the capital cost of \$3.7B for one specific C$^3$-250 scenario in~\cite{Bai:2021rdg} the
 XCC represents a savings of \$1.4B.  Given the very early stage of the XCC design and the many XFEL technical challenges, it is important that Table~\ref{tab:capitalcost} be viewed as illustrative, providing insight into the {\it potential} cost savings of the XCC.

 \begin{table}

    \centering
    \begin{tabular}{    c  | c | c | c} \toprule
    & Sub-Domain & \%  & \% \\ \midrule
    Sources                & Injectors                      & 9  & 26 \\
                           & FEL                            & 9  &    \\
                           & Beam Transport                 & 9  &    \\ \midrule
 Main Linac                & Cryomodule                     & 9  & 30   \\
                           &  C-band Klystron               & 22  &     \\  \midrule
    BDS                    &  Beam Delivery and Final Focus &  7  &  15  \\
                           &         IR                     &  8  &    \\ \midrule
Support Infrastructure     & Civil Engineering              &  5  &  28  \\
                           &  Common Facilities             & 18  &    \\
                           &  Cryo-plant                    &  6  &    \\ \midrule 
        Total              &      2.3B\$                    & 100 & 100   \\

        \bottomrule
    \end{tabular}
        \caption{ \label{tab:capitalcost} Initial estimate of  XCC cost breakout using the C$^3$-250 cost model.}
 \end{table}

\newpage

\section{Design Overview}
\subsection{Attainable Energy}

\subsubsection{Electron Accelerator} \label{accelerator}

The C$^3$ technology represents a new methodology for dramatically reducing the cost of high gradient accelerators, while increasing their capabilities in terms of gradient and efficiency. After two decades of exploring the high gradient phenomena observed in room-temperature accelerator structures, the underlying physics models related to these phenomena have been deduced. This knowledge led  to the creation of a new paradigm for the design of accelerator structures, which includes: a new topology for the structure geometry \cite{Sami1,Sami2} operating at cryogenic temperature, the use of doped copper in the construction of these structures \cite{Sami4}, and a new methodology for the selection of operating frequency bands \cite{Sami4}. In particular, for science discovery machines, optimization exercises have revealed that the optimal frequency should be around 6--8 GHz for operation with a gradient well above 100 MeV/m while maintaining exquisite beam parameters.  That explains why both UCLA and LANL are trying very hard to build their infrastructure at C-band (5.712 GHz), a frequency band that is close enough to the optimal point, but with some industrial support behind it. 

Furthermore, the so-called ``distributed-coupling structure" \cite{Sami1} and its operation at cryogenic temperature represent a breakthrough for the e$^-$ source. Electron guns can be designed around this concept with an unprecedented brightness  \cite{Sami6}. Using this technology can result in an extremely economical system for this $\gamma$-$\gamma$ collider. The two Linacs required for the collider could be made extremely compact due to the high gradient capabilities of the C$^3$ technology and the limited energy reach required of 62.5 GeV. With the bright electron beam sources, damping rings can be eliminated.   The Linac parameter set is included in Tables~\ref{tab:design} and~\ref{tab:designeminus}.  A description of C$^3$ technology applied to an $e^+e^-$ collider can be found here\cite{Bai:2021rdg}.

\subsubsection{X-ray FEL }

The two identical X-ray FEL lines, which provide the necessary circularly-polarized 1.2 nm (1 keV) photons, can be constructed using a long helical undulator. A full time-dependent GENESIS study (described below) has validated the high magnetic field and high electron energy design considered here.  The quantum diffusion energy spread in such an undulator must be taken into account, and will be properly included in future studies. As the main Linac can accelerate electrons to 62.5 GeV, we take the electron energy for the XFEL line to be around 31 GeV, with normalized emittance of 120 nm, bunch charge of 1 nC and relative RMS slice energy spread of $\langle\Delta\gamma/\gamma\rangle$ of 0.05\%. 

Using a permanent-magnet undulator, with peak magnetic field slightly above 1 Tesla, undulator period around 9 cm and an average $\beta$-function of 12 m, we can produce 1 keV X-ray pulse energy $\sim 0.07$ J at FEL saturation length of roughly 60 m and with negligible quantum diffusion effects~\cite{KimHuangBook,HuangReviewPaper}. As we know from a decade of X-ray FEL studies, if we can produce a seeded FEL (such as through self-seeding or other similar processes) and taper the undulator's $K$ parameter after saturation, we can continue to extract X-ray pulse energy with an order-of-magnitude improvement in efficiency ~\cite{TaperPaper}. Then we can reach the targeted pulse energy of 0.7 J at 1 keV photon energy, which is about 2.3\% of the electron beam energy. The overall length of the undulators is estimated to be within 200 m. This is just an example parameter set (summarized in Table~\ref{tab:design} below). 

\begin{figure}
\centering
     \includegraphics[width=0.75\textwidth]{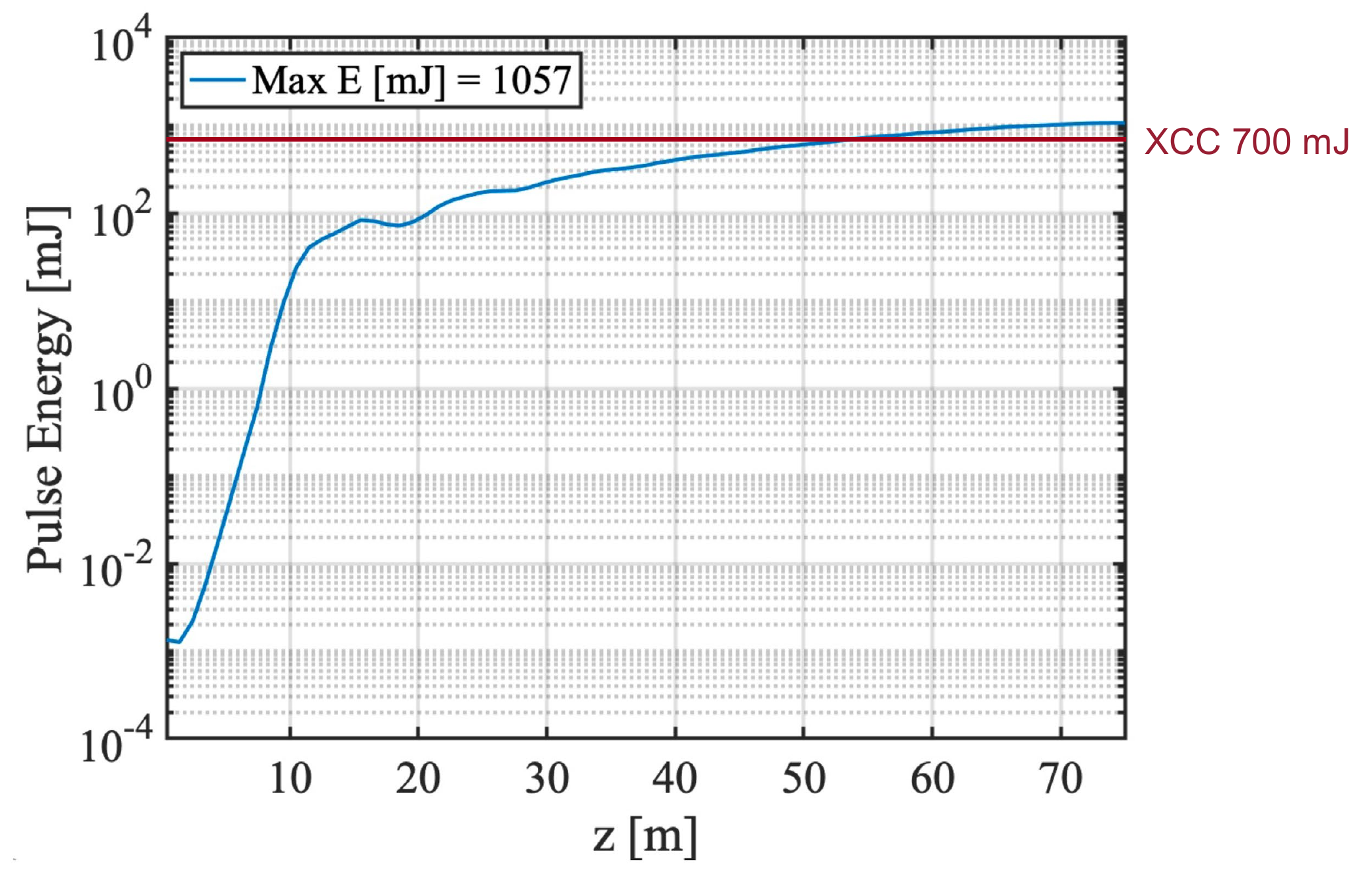}
     \caption{GENESIS simulation of X-ray pulse energy versus distance along the XCC XFEL undulator.}
     \label{fig:pulseevsz}
 \end{figure}

The XFEL design has been simulated with GENESIS 1.3. The simulation is run with parameters similar to those shown in table 2. The difference between those and the simulated parameters are the initial electron slice energy spread (0.01$\%$), electron energy (30 GeV), peak current (6kA), average beta function (2 m), undulator period (5 cm) and undulator peak field (1.8 T). The undulator is a helical permanent magnet undulator with a super-imposed alternating gradient focusing quadrupole lattice. The undulator is split into two sections, the first one generates Self Amplified Spontaneous Emission (SASE) \cite{Bonifacio1984} and the second is a self-seeded section in which a monochromatic seed is overlapped with the electron beam after passing through an idealized monochromator. The time dependent SASE simulation produces a 0.4 mJ and $\sim$ 100 fs FWHM X-ray pulse after a 15 m undulator. The self-seeded section assumes 0.3 $\%$ of the SASE X-ray power is filtered through the monochromator and a quadratic post-saturation taper is applied to increase the pulse energy following the exponential gain region. As shown in Fig.~\ref{fig:pulseevsz}, the resulting X-ray energy after a 75 m undulator is 1.05 J with a 3.5 $\%$ extraction efficiency, in reasonable agreement with the analytic estimates obtained using the parameters of Table 2. We note that a similar extraction efficiency has already been achieved experimentally for tapered self-seeded systems XFELs \cite{Emma2017}.

%====== summary table ========

\begin{table}[h]
\begin{center}
\caption{\label{tab:design} Summary of design parameters for $\gamma\gamma$ mode at $\sqrt{s}=125$~GeV.}
\begin{tabular}{ |l|l|||l|l|c| }
\hline
Final Focus parameters  & Approx.~value & XFEL parameters & Approx.~value \\
\hline
\hline
Electron energy & 62.8 GeV & Electron energy & 31 GeV \\ 
Electron beam power & 0.57 MW & Electron beam power &  0.28 MW \\
$\beta_x/\beta_y$ & 0.03/0.03 mm & normalized emittance & 120 nm \\ 
$\gamma\epsilon_x/\gamma\epsilon_y$ & 120/120 nm & RMS energy spread $\langle\Delta\gamma/\gamma\rangle$ &  0.05\% \\ 
$\sigma_x/\sigma_y$ at $e^-e^-$ IP & 5.4/5.4 nm & bunch charge & 1 nC \\ 
$\sigma_z$ & 20 $\mu\rm{m}$  & Undulator B field & $\gtrsim$ 1 T \\ 
bunch charge & 1 nC & Undulator period $\lambda_u$ &  9 cm \\ 
Rep. Rate at IP & $240\times38$ Hz & Average $\beta$ function &  12 m \\ 
$\sigma_x/\sigma_y$ at IPC & 12.1/12.12 nm & x-ray $\lambda$ (energy) &  1.2 nm (1 keV)\\ 
$\mathcal{L}_\textrm{geometric}$ & $9.7\times 10^{34}\ \textrm{cm}^2\ \textrm{s}^{-1}$ & x-ray pulse energy &  0.7 J \\ 
$\delta_E/E$ & 0.05\% & pulse length &  40 $\mu\rm{m}$   \\
$L^*$ (QD0 exit to $e^-$ IP) &  1.5m  & $a_{\gamma x}$/$a_{\gamma y}$ (x/y waist) & 21.2/21.2~nm \\
$d_{cp}$ (IPC to IP) & $60\ \mu$m  & non-linear QED $\xi^2$ & 0.10 \\
%$\eta'^*_x$ & 20 mrad &  & \\
QD0 aperture & 9 cm diameter & & \\  \midrule
Site parameters & Approx. value & & \\ \midrule
crossing angle & 2 mrad & & \\
total site power & 86 MW & & \\
total length & 2.5 km & & \\
\hline
\end{tabular}
\end{center}
\end{table}

\begin{table}[h]
\begin{center}
\caption{\label{tab:designeminus} Summary of design parameters for $e^- \gamma$ mode at $\sqrt{s}=140$~GeV.}
\begin{tabular}{ |l|l|||l|l|c| }
\hline
Final Focus parameters  & Approx.~value & XFEL parameters & Approx.~value \\
\hline
\hline
Electron energy & 70.0 GeV & Electron energy & 31 GeV \\ 
Electron beam power & 0.64 MW & Electron beam power &  0.28 MW \\
$\beta_x/\beta_y$ & 0.03/0.03 mm & normalized emittance & 120 nm \\ 
$\gamma\epsilon_x/\gamma\epsilon_y$ & 1200/12 nm & RMS energy spread $\langle\Delta\gamma/\gamma\rangle$ &  0.05\% \\ 
$\sigma_x/\sigma_y$ at $e^-e^-$ IP & 16.2/1.6 nm & bunch charge & 1 nC \\ 
$\sigma_z$ & 10 $\mu\rm{m}$  & Undulator B field & $\gtrsim$ 1 T \\ 
bunch charge & 1 nC & Undulator period $\lambda_u$ &  9 cm \\ 
Rep. Rate at IP & $240\times38$ Hz & Average $\beta$ function &  12 m \\ 
$\sigma_x/\sigma_y$ at IPC & 17.1/1.71 nm & x-ray $\lambda$ (energy) &  1.2 nm (1 keV)\\ 
$\mathcal{L}_\textrm{geometric}$ & $1.1\times 10^{35}\ \textrm{cm}^2\ \textrm{s}^{-1}$ & x-ray pulse energy &  0.7 J \\ 
$\delta_E/E$ & 0.05\% & pulse length &  40 $\mu\rm{m}$\\
$L^*$ (QD0 exit to $e^-$ IP) &  1.5m  & $a_{\gamma x}$/$a_{\gamma y}$ (x/y waist) & 15.3/10.0~nm \\
$d_{cp}$ (IPC to IP) & $10\ \mu$m  &  non-linear QED $\xi^2$ & 0.29 \\
%$\eta'^*_x$ & 20 mrad &  & \\
QD0 aperture & 9 cm diameter & & \\ \midrule
Site parameters & Approx. value & & \\ \midrule

crossing angle & 2 mrad & & \\
total site power & 90 MW & & \\
total length & $\sim 2.5$ km & & \\
\hline
\end{tabular}
\end{center}
\end{table}

\subsection{Attainable Luminosity}

\subsubsection{Electron Final Focus}

\begin{figure}
\centering
     \includegraphics[width=0.95\textwidth]{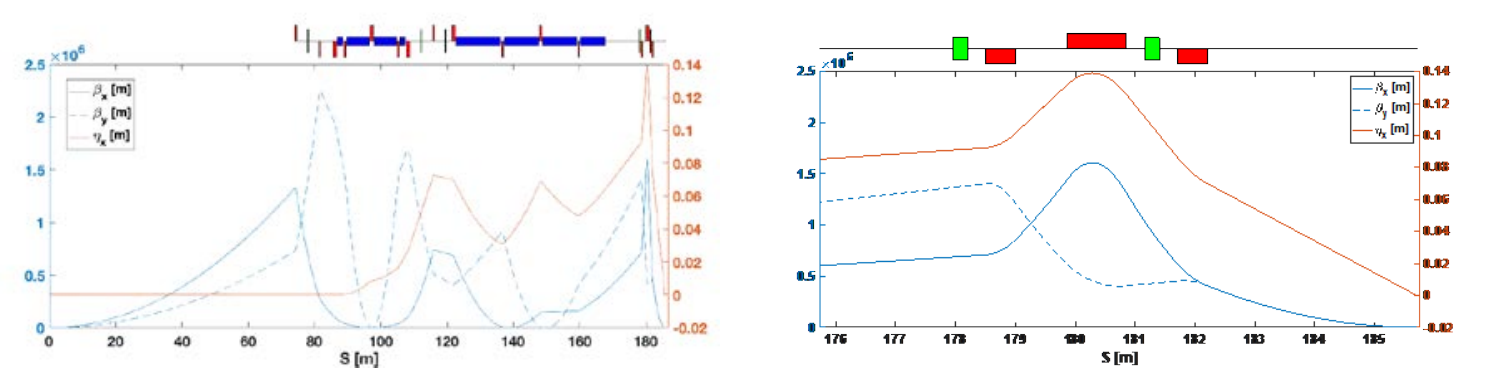}
     \caption{Beta and horizontal dispersion functions for final focus beamline (IP is on the right). The right figure shows a close up of the IP triplet.}
     \label{fig:betadisp}
 \end{figure}

A preliminary layout for the final focus system is shown in Fig.~\ref{fig:betadisp}.
This is not an optimized design at this stage and is shown for illustration purposes only. 
The length of the system is about 110m as shown, using realistic magnet strengths. The design follows 
that used for ILC and CLIC, namely the local chromatic compensation scheme proposed by Raimondi \& Seryi\cite{Raimondi:2000cx} and tested at ATF2, KEK\cite{White:2014vwa}. The design uses a pair 
of sextupole magnets located locally to the final triplet to cancel the chromaticity generated by the final focus system magnets. An interleaved, second pair of sextupoles are 
used to simultaneously cancel geometric aberrations introduced by the chromatic correction sextupoles, with a fifth sextupole 
used to help control third-order aberrations generated by the interleaved sextupole pairs. Octupoles, decupoles (and perhaps higher harmonic magnets) will also be required but are not 
shown here. Horizontal dispersion, required for the chromatic cancelation to occur, is generated by three families of 
bend magnets. The length of the bend magnets is chosen such 
that negligible emittance growth due to synchrotron radiation exists.

 \subsubsection{X-ray Optics}\label{xrayoptics}
 The x-ray beam must be focused from a waist radius of $a_\gamma=9000$~nm at the undulator exit to $a_\gamma=30$~nm (70~nm FWHM) at the Compton interaction point (IPC) in order to match the  transverse size of the electron beam at the IPC.  The distance $d_{cp}=60\ \mu$m between the IPC and IP cannot be made much larger due to the angular spread of the converted 62.5 GeV photons.   The angle $\theta$
 that a converted photon makes with respect to the electron direction is correlated with its energy $\omega$ via $\theta=\theta_0\sqrt{\omega_m/\omega-1}$, where $\omega_m=62.8$~GeV is the maximum photon energy and $\theta_0=\sqrt{x+1}/\gamma_e=0.26$~mrad.  A photon 
 with energy $m_H/2=62.5$~GeV therefore makes an angle $\theta=0.018$~mrad,
 so that the design value $d_{cp}=60\ \mu$m
 results in a 1.1~nm increase in the transverse size of the 62.5~GeV photon beam.  This should be compared to the 5.4~nm transverse size of the photon beam at the IP assuming no Compton angular spread.
 As an example of an alternate configuration, if the distance between the IPC and IP  were increased to $d_{cp}$=100~nm, the $\gamma\gamma$ luminosity would drop by 30\% with respect to the luminosity with $d_{cp}=60\ \mu$m.
 
 The x-ray optics design is still under development.  There are some major challenges for implementing the high numerical aperture (NA) requirement for the beam focusing in the soft x-ray regime where traditional designs employ small NA grazing incidence setups, sometimes using several stages of focusing.  In these cases, the focal length of the upstream mirror can exceed the length of the downstream mirror, which can limit the NA of the system, so while grazing incidence is preferred from the point of view of throughput and minimization of the thermal deformation and radiation damage, the limited NA is incompatible with the high focusing requirement for this application.  In contrast, recent ultrashort focal length KB mirror systems that have higher NA with expected focusing spot sizes of 39 nm FWHM at 1 keV\cite{10.1117/12.2569845} have been proposed.  However, these mirrors may be more prone to thermal deformation and radiation damage.  
 
\begin{figure}
\centering
     \includegraphics[width=0.75\textwidth]{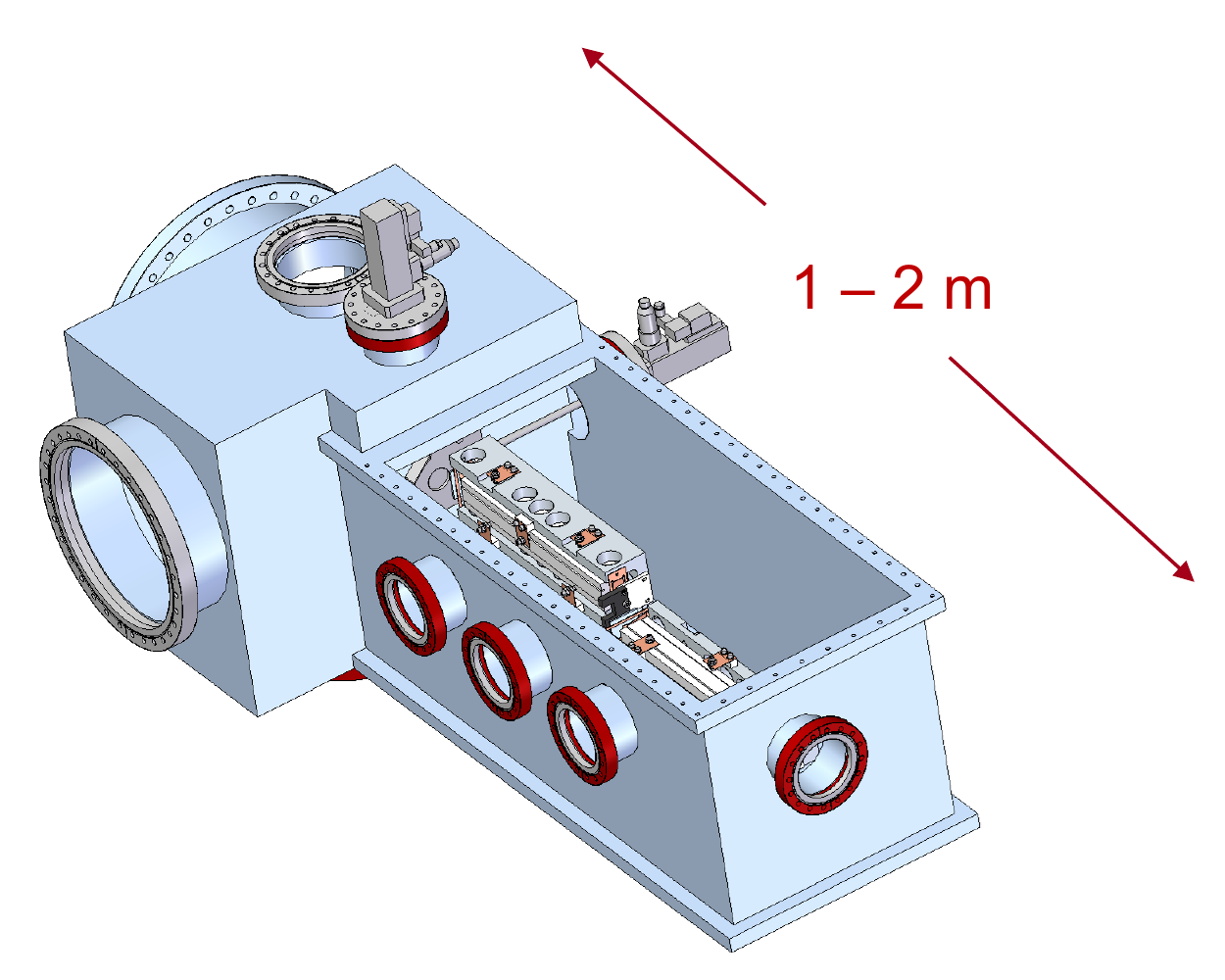}
     \caption{Representative LCLS KB mirror chamber with vertical and horizontal KB mirrors, movers, and cooling hardware.}
     \label{fig:kblcls}
 \end{figure}

\begin{figure}
\centering
     \includegraphics[width=0.85\textwidth]{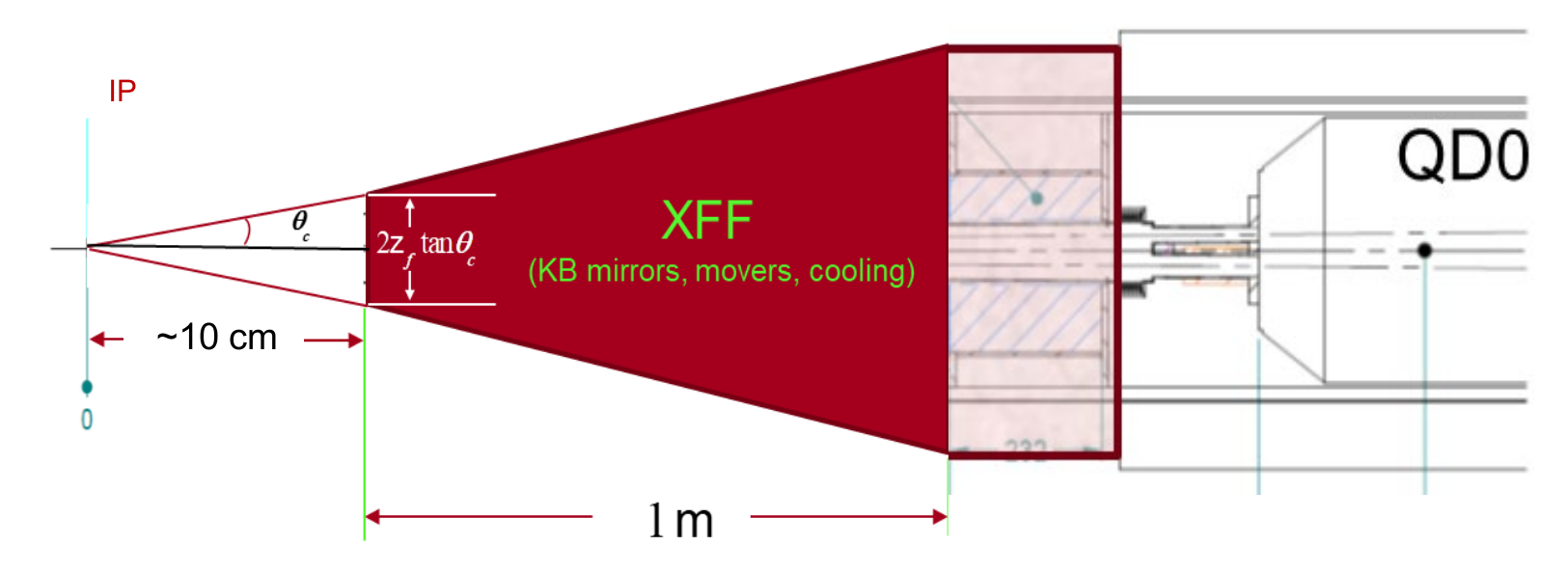}
     \caption{X-ray final focus chamber (XFF) superimposed on the SiD beamline at ILC.  The
     pink-shaded region between the XFF and QD0  region can't be instrumented due to the XFF material, and can therefore be utilized either  to extend the XFF backwards (if, for example, more space is needed for the mirror cooling system) or to move QD0 closer to the IP.}
     \label{fig:xffschematic}
 \end{figure}

 While this is a challenge in the soft x-ray regime, multi-stage high NA Kirkpatrick-Baez (KB) mirrors have been built for hard x-rays, for example achieving a 25~nm FWHM spot size at 15 keV\cite{Yamada:19}.  The x-ray beam in this case can be focused in several stages using Kirkpatrick-Baez (KB) mirrors under grazing incidence (small angle of incidence). Each stage consists of a vacuum chamber containing two KB mirrors (one each for horizontal and vertical focusing), movers, and cooling hardware. An example of an LCLS KB mirror chamber is shown in  Fig.~\ref{fig:kblcls} After passing through all but the final focusing stage, the x-ray beam enters the electron vacuum pipe just after the last FF bend.    About 1~meter from the IP the x-ray beam passes through the final x-ray focusing chamber  (XFF) as shown in  Fig.~\ref{fig:xffschematic}.  %The XFF is conically shaped and centered on the electron beam line.  
 The angle $\theta_c$ subtended by the XFF cone represents a dead region for the detector.  Based on Higgs physics requirements, the target design is $\theta_c<200$~mrad.  Further optics research is needed to determine which values of $\theta_c$ can be achieved in practice and what the ideal mirror geometry will be for 1 keV soft x-rays.
 
%   However, for x-ray energies less than 4 keV, KB mirrors systems present challenges due to their long focal lengths.  
% {\color{red} (Should explain why the long focal lengths are a problem for energies $< 4$~keV. In particular, why --- given that ${\rm NA}\approx \theta_c \propto \lambda$ --- is the following relation violated for photon energies less than about 4~keV: ${\rm min}(a_\gamma) \approx \lambda/{\rm NA}\approx\lambda/\theta_c\approx 10$~nm independent of $\lambda$?   See section 3.3.1 of~\cite{Salditt2020})} 
 
 \subsubsection{Beam-beam Effects and Luminosity Integral}
 
 The CAIN Monte Carlo program is used to simulate beam-beam effects at both the IPC and IP. For  laser scattering at the IPC, the  CAIN program includes non-linear QED effects in Compton scattering ($e^-\gamma_0\rightarrow e^-\gamma$) and Breit-Wheeler scattering ($\gamma\gamma_0\rightarrow e^+e^-$) where $\gamma$ and $\gamma_0$ refer to the Compton-scattered and laser photon, respectively.  Bethe-Heitler scattering with laser photons ($e^-\gamma_0\rightarrow e^-e^+e^-$) is not included in CAIN, and must be added to the code.  To include this process, the equivalent photon approximation is used for virtual photons $\gamma^*$ radiated by the initial state electron,
 and the code for $\gamma\gamma_0\rightarrow e^+e^-$ is then used to simulate $\gamma^*\gamma_0\rightarrow e^+e^-$.  Non-linear QED Bethe-Heitler scattering with laser photons is not simulated.

 In the nominal XCC polarization configuration of $2P_c\lambda_e=+0.9$, the electron and laser photon helicities are given the same sign, which leads to collision lengths of 34, 25, and 95~$\mu\rm{m}$ for the Compton process
$e^-\gamma_0\rightarrow e^-\gamma$, the trident process $e^-\gamma_0\rightarrow e^-e^+e^-$,
and the  $\gamma\gamma_0$ annihilation process $\gamma\gamma_0\rightarrow e^+e^-$, respectively.      The $\gamma\gamma_0$ annihilation collision length is $3\times$  longer than it would have been if the electron and laser beams were unpolarized, and
$5\times$ longer than the collision length if the electron and laser beams had been given opposite helicities. With a collision length of 6.3 times the total laser pulse length, the $\gamma\gamma_0$ annihilation process is a nonissue.
The total conversion efficiency of electrons to primary first generation photons is 18\%.  

\begin{figure}
\centering
     \includegraphics[width=0.75\textwidth]{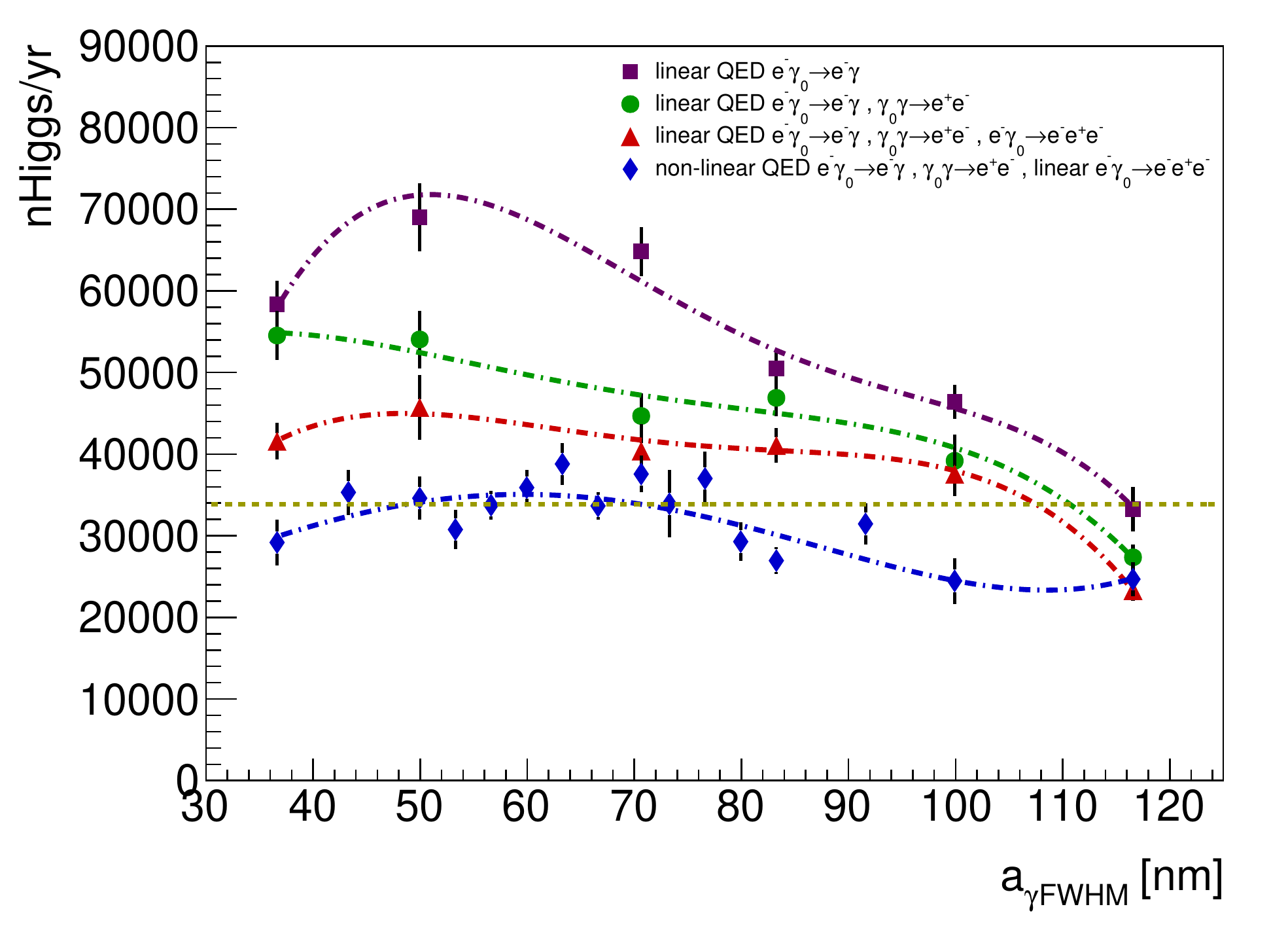}
     \caption{Scans of the FWHM spot size ${\rm a}_{\gamma FWHM}$ of the XFEL beam at the Compton interaction point (IPC) showing the impact of different IPC processes on the Higgs production rate. In the expressions for the IPC processes the symbols 
     $\gamma_0$ and $\gamma$ refer to laser photon and scattered photon, respectively. 4th order polynomial fits to the CAIN simulation results are indicated by dashed lines. The gold horizontal dashed line corresponds to the XCC design Higgs production rate.}
     \label{fig:lumVsAgamma}
 \end{figure}

\begin{figure}
\centering
     \includegraphics[width=0.75\textwidth]{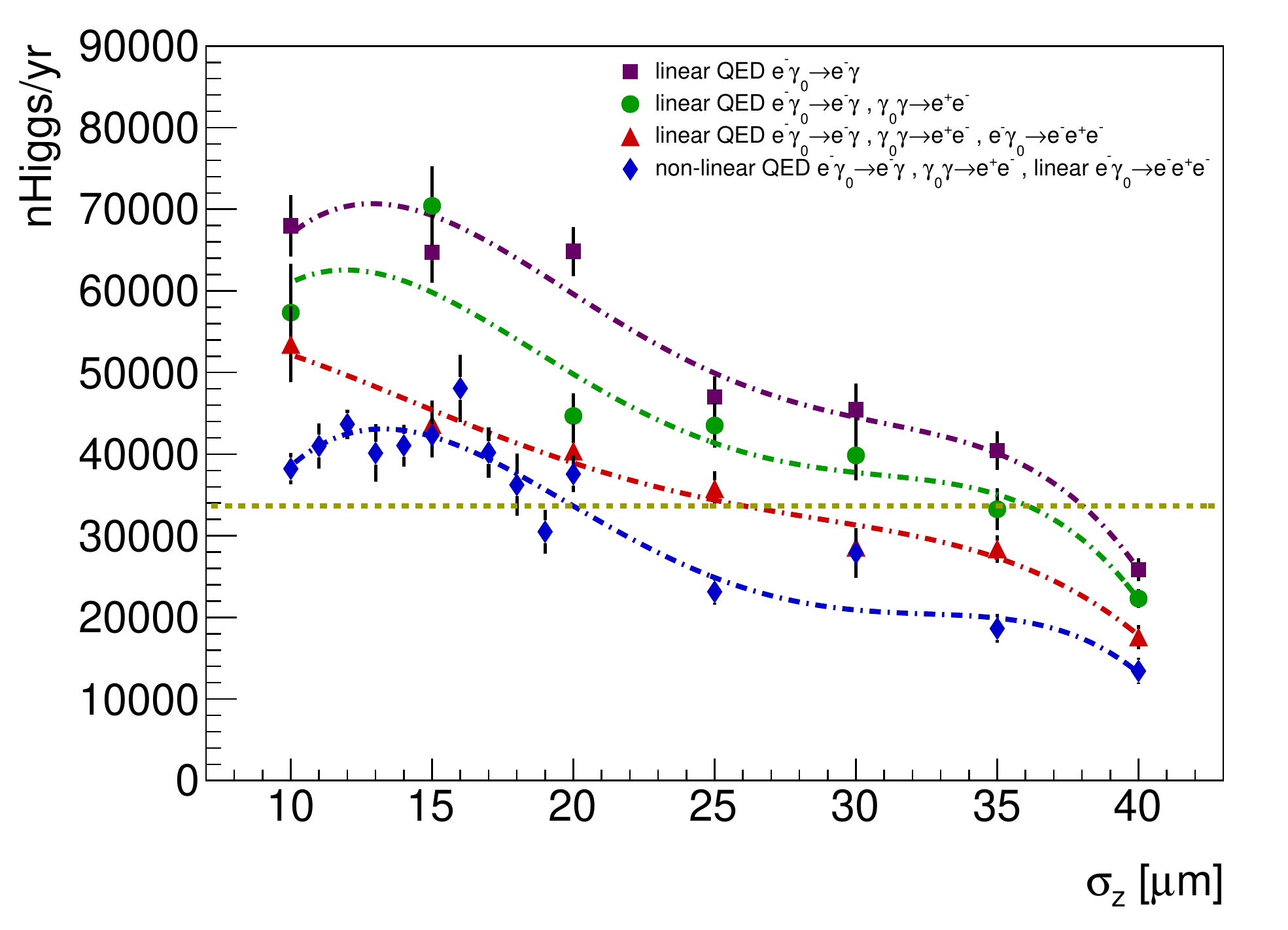}
     \caption{Scans of the r.m.s. electron longitudinal  bunch density $\sigma_{ez}$ showing the impact of different IPC processes on the Higgs production rate. In the expressions for the IPC processes the symbols 
     $\gamma_0$ and $\gamma$ refer to laser photon and scattered photon, respectively.  4th order polynomial fits to the CAIN simulation results are indicated by dashed lines. The gold horizontal dashed line corresponds to the XCC design Higgs production rate.}
     \label{fig:lumVsSigez}
 \end{figure}

\begin{figure}
\centering
     \includegraphics[width=0.75\textwidth]{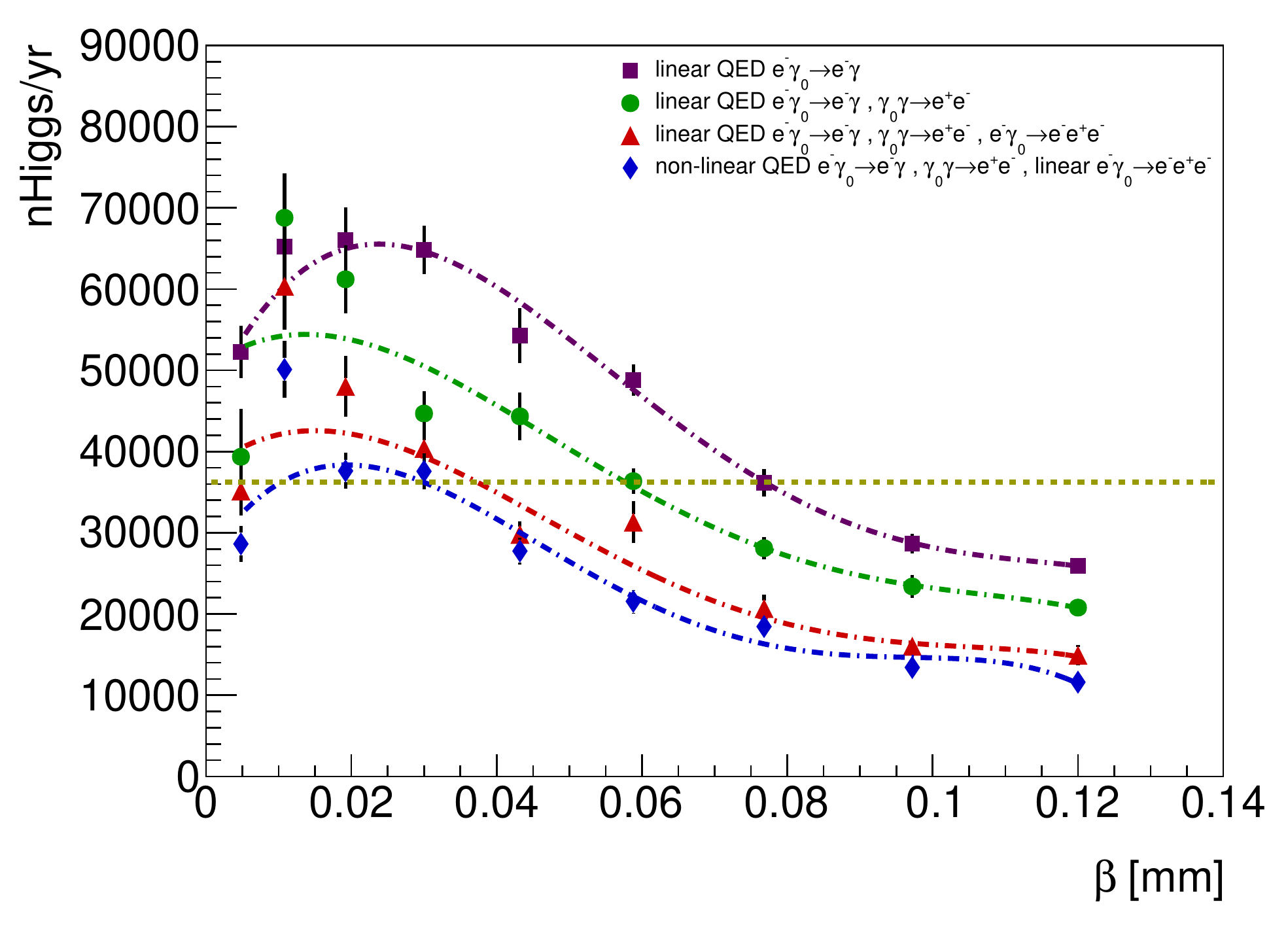}
     \caption{Scans of the electron final focus $\beta$ value showing the impact of different IPC processes on the Higgs production rate. In the expressions for the IPC processes the symbols 
     $\gamma_0$ and $\gamma$ refer to laser photon and scattered photon, respectively.  4th order polynomial fits to the CAIN simulation results are indicated by dashed lines. The gold horizontal dashed line corresponds to the XCC design Higgs production rate.}
     \label{fig:lumVsBeta}
 \end{figure}

Scans of the XFEL FWHM spot size,  r.m.s. electron longitudinal bunch density and electron final focus $\beta$ value are shown in Figs.~\ref{fig:lumVsAgamma},~\ref{fig:lumVsSigez}, and \ref{fig:lumVsBeta}, respectively.  In addition to illustrating the sensitivity of the luminosity to laser waist, electron bunch length and electron spot size, the plots demonstrate the impact of different Compton processes on the luminosity.  Note the large effect from the Bethe-Heitler process $e^- \gamma_0 \rightarrow e^- e^+ e^-$ in Figs.~\ref{fig:lumVsSigez} and \ref{fig:lumVsBeta}. 

Figs.~\ref{fig:lumVsAgamma} illustrates an interesting interplay between different effects as the photon density is increased.  As the photon density is increased
(i.e., ${\rm a}_{\gamma FWHM}$ is made smaller) the Compton conversion efficiency increases but the rate for
Bethe-Heitler production $e^-\gamma_0\rightarrow e^-e^+e^-$ and the non-linear
QED parameter $\xi^2$ also increase.  The result is a flat distribution in the Higgs production rate versus ${\rm a}_{\gamma FWHM}$ when all effects are include (blue curve).

At the IP the CAIN program simulates beamstrahlung, coherent pair-production, and the incoherent particle-particle interactions $\gamma\gamma\rightarrow e^+e^-$, $\gamma e^\pm \rightarrow e^\pm e^+e^-$, and $ee\rightarrow ee e^+e^-$.   For $\gamma \gamma$ collisions, the most important process is beamstrahlung, which produces the large low-side tail in luminosity for $0<E_{\gamma\gamma}<100$~GeV  seen in Figs.~\ref{fig:x1000lumiplus} and \ref{fig:x1000lumiminus}.  For $e^- \gamma$ collisions coherent pair-production also plays a major role as the positrons from the pair-production start to focus the opposing electron beam to a smaller size.  This will be discussed in more detail in Sec.~\ref{egammachallenge}.

\begin{figure}
\centering
     \includegraphics[width=0.95\textwidth]{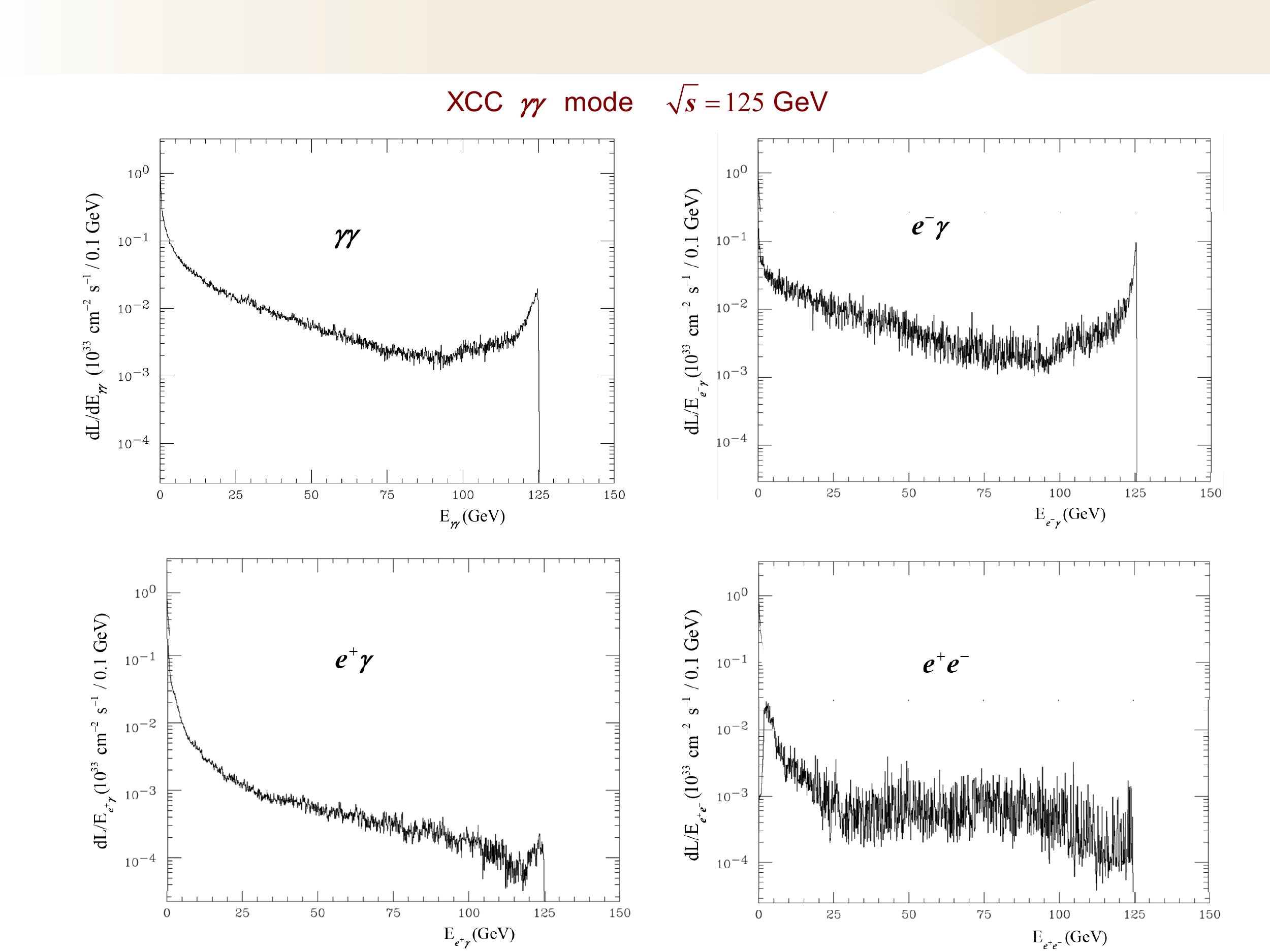}
     \caption{Luminosity spectra for different processes for XCC in the $\gamma\gamma$ mode at $\sqrt{s}=125$~GeV.}
     \label{fig:allfour125}
 \end{figure}

The luminosity for various processes in the $\gamma\gamma$ mode of XCC at $\sqrt{s}=125$~GeV as calculated by CAIN was shown in Table~\ref{tab:lumisummary}.  The luminosity distributions for four of the processes is in the $\gamma\gamma$ mode shown in Fig.\ref{fig:allfour125}.

 The luminosity for various processes in the $e^- \gamma$ mode of XCC at $\sqrt{s}=140$~GeV as calculated by CAIN is shown in Table~\ref{tab:lumieminussummary}.  
 %The luminosity distributions for four of the processes in the $e^-\gamma$ mode is shown in Fig.\ref{fig:allfour140}.

 \begin{table}

    \centering
    \begin{tabular}{    c  | c | c  } \toprule
        \multicolumn{3}{ c}{$e^-\gamma$ mode $\sqrt{s}=140$~GeV} \\ \midrule
       & \multicolumn{2}{ c}{Luminosity ($10^{34}$ cm$^{-2}$ s$^{-1}$)} \\[3pt]
        Process  & Total &   $\sqrt{\hat{s}}>139$~GeV \\ \midrule
     $e^-\gamma$     & 11.5 & 0.32 \\
     $\gamma\gamma$  & 14.5 &  -  \\
     $e^-e^+$        & 13.4 &  - \\
     $\gamma e^+$     & 11.3 &  -  \\
     $\gamma e^-$     & 0.92 &   - \\
     $e^-e^-$        & 0.43 & 0.07 \\
     $e^+\gamma$     & 0.09 &  -  \\
     $e^+e^-$        & 0.01 & - \\  
     
        \bottomrule
    \end{tabular}
        \caption{ \label{tab:lumieminussummary}Total luminosity and luminosity for  $\sqrt{\hat{s}}>139$~GeV for different processes at the XCC running in $e^-\gamma$ mode at $\sqrt{s}=140$~GeV.}
\end{table}

 \subsubsection{Beam Extraction}
 
\begin{figure}
\centering
     \includegraphics[width=0.85\textwidth]{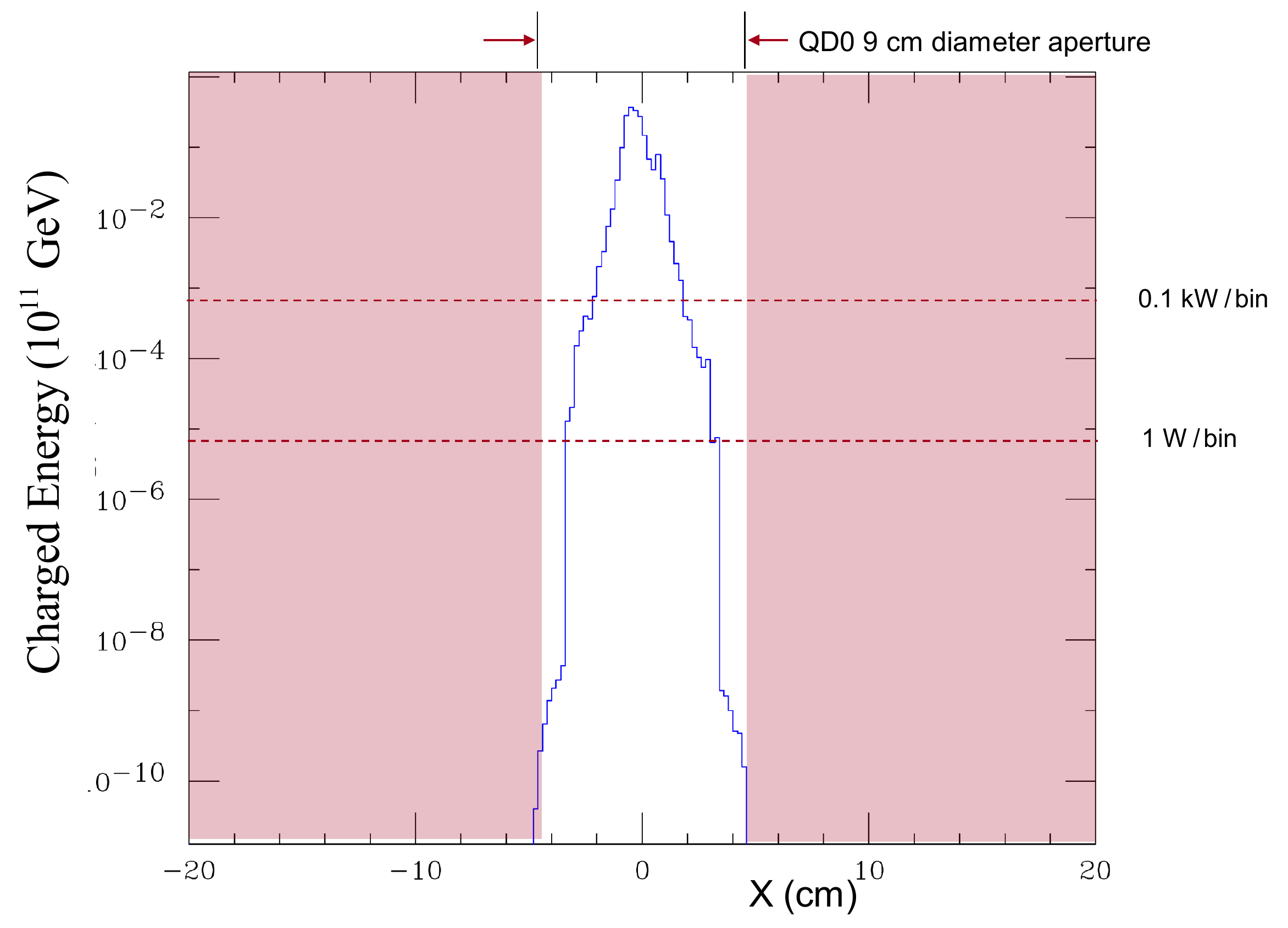}
     \caption{ Charged energy per beam per bunch crossing versus position transverse to the beam (X) at the distance $L^*=1.5$~m downstream of the IP.}
     \label{fig:evsx150}
 \end{figure}

\begin{figure}
\centering
     \includegraphics[width=0.85\textwidth]{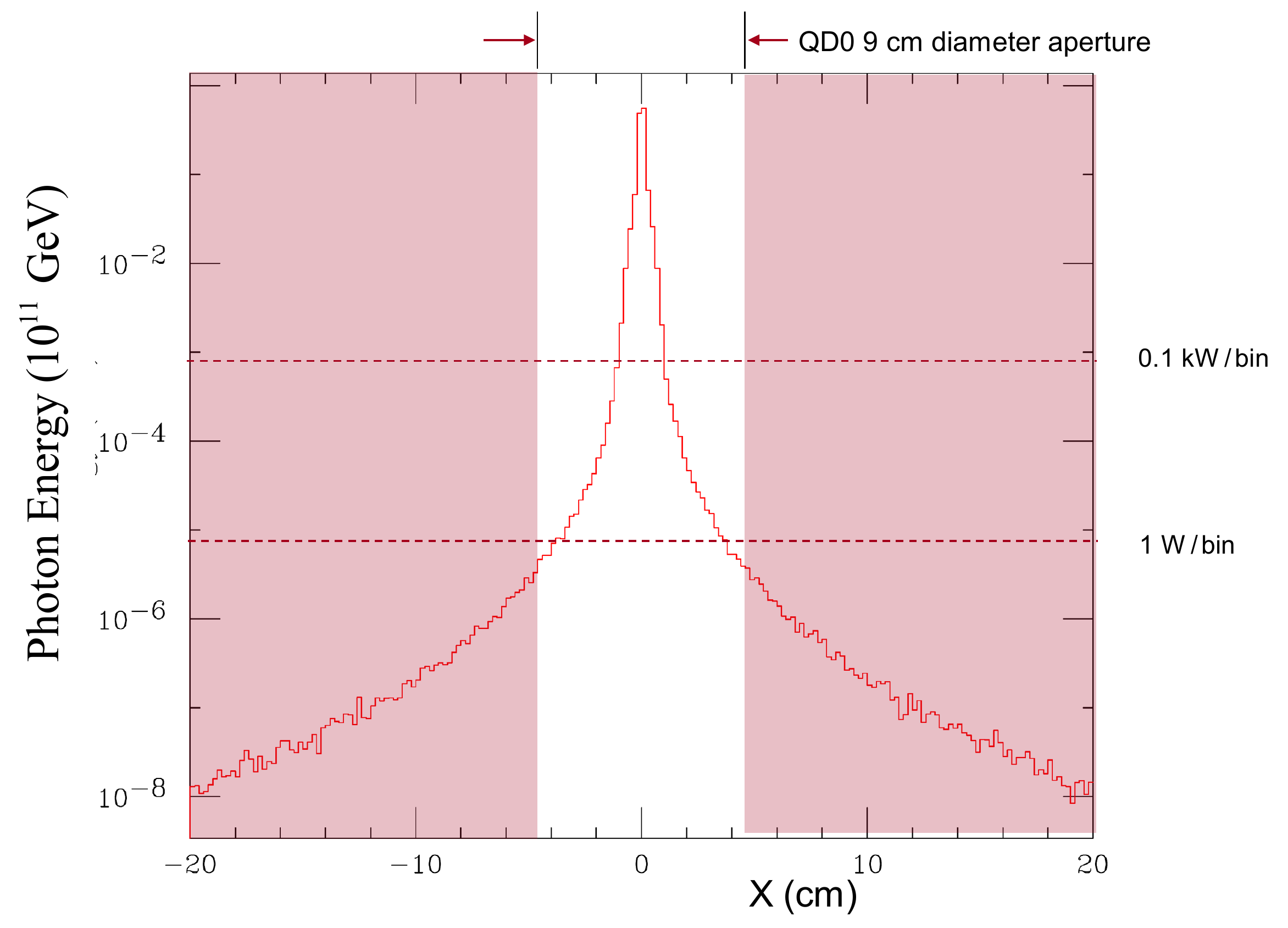}
     \caption{ Neutral energy per beam per bunch crossing versus position transverse to the beam (X) at the distance $L^*=1.5$~m downstream of the IP.}
     \label{fig:ephvsx150}
 \end{figure}

 The charged and neutral energy profiles several meters downstream of the IP, along with the strong anticorrelation between luminosity and crossing angles at both the IP and IPC, favor a nearly head on collision.  A small crossing angle of 2~mrad was considered
 at one point for the ILC~\cite{Nosochkov:2005ip}, and this will serve as the starting point for the XCC beam extraction design.  The QD0 aperture in~\cite{Nosochkov:2005ip} was 9~cm in diameter, which is the baseline aperture for QD0 in the XCC.   Figs.~\ref{fig:evsx150} and
 \ref{fig:ephvsx150} show the energy deposition per beam per bunch crossing as a function of position transverse to the beam at the distance $L^*=1.5$~m downstream of the IP.  In stable operation the energy deposition on QD0 is a few Watts, assuming no masking between the IP and QD0.  The energy of the photons striking QD0 in the region $|X|>4.5$~cm is about 1~MeV. 
 
 \newpage
 
\section{Design Challenges}
\subsection{Electron Final Focus}
The  final focus design differs from than that of ILC \& CLIC in 2 key aspects, each of which raises concerns which need further studies to address:
\begin{itemize}
    \item 	Round beams at the IP leads to the preference of a final triplet instead of final doublet configuration. The required angular dispersion at the IP is about double that required for ILC/CLIC to achieve the same dispersion at the sextupole locations. This will have an adverse effect on the momentum acceptance of the extraction line and may lead to increased detector backgrounds.
	\item	The requested IP beta functions are 0.03 mm in both planes. This should be compared with 11 x 0.48 mm for the baseline ILC design (and a corresponding design tested at the ATF2 facility). The much smaller $\beta^*$ values here generate significantly higher chromatic distortions, requiring stronger sextupole corrections. This in turn requires more finely tuned 3rd, 4th+ order corrections to compensate for the sextupoles. Experience from CLIC tuning studies and operational experience at ATF2 has shown that tolerances become rapidly tighter (magnet field quality and positional tolerances) as $\beta^*$  is lowered below ILC values, and online tuning becomes harder and takes longer. Also, operational experience at ATF2 showed that tuning becomes more difficult with smaller $\beta^*_x$  : $\beta^*_y$ ratios (where the smallest spot sizes were only accomplished at 10X design $\beta^*_x$). With this in mind, a careful study of the tolerances of this, modified, final focus design is important to understand the ramifications on expected delivered luminosity.
	\end{itemize}
	
 \subsection{X-ray Optics}
 As discussed in section~\ref{xrayoptics}, the 1 keV X-ray spot size of 70~nm FWHM is a demanding specification. One recent ultrashort KB mirror design that combines a short focal length with total reflection predicts that a 39~nm spot size at 1 keV is possible, but this configuration will present significant challenges in being able to handle the 700 mJ pulse energy and a $38\times 240$~Hz repetition rate of the XCC. Thermal deformation and radiation damage  can result in focused beam distortion and degrade mirror and/or coating performance. While active research is currently being performed in cryo-cooling of mirrors to alleviate some of the degradation, challenges such as mechanical vibration caused by the cooling will need to be overcome.  Certain materials such as SiC are known to have higher radiation tolerance. However, growing these materials to the footprint required for an x-ray mirror and developing polishing methods remain a challenge. Further research into mirror design and composition to address these challenges will be required.

 \subsection{$e^-\gamma$ Luminosity at 140 GeV}~\label{egammachallenge}
 In the running scenario described in Sec~\ref{sec:higgsphysics}, two years are spent collecting $e^- \gamma\rightarrow e^-H$ events at $\sqrt{s}=140$~GeV for every year collecting $\gamma\gamma\rightarrow H$ events at $\sqrt{s}=125$~GeV.  With the baseline design, the Higgs rate in $e^- \gamma$ collisions is 0.8\% of the Higgs rate in $\gamma\gamma$ collisions at $\sqrt{s}=125$~GeV.  This is an unsatisfactory situation as two-thirds of the running time is spent waiting for a small number of $e^- \gamma\rightarrow e^-H$ events to dribble in.
 
 The XCC physics program would be enhanced if the $e^- \gamma$ luminosity could be 
 significantly increased.  From Table~\ref{tab:designeminus} the $e^-e^-$ geometric luminosity is    $1.1\times 10^{35}\ \textrm{cm}^2\ \textrm{s}^{-1}$, and yet from Table~\ref{tab:lumieminussummary} the useful luminosity for $e^- \gamma$ collisions with $\sqrt{\hat{s}}>139$~GeV is only $3.2\times 10^{33}\ \textrm{cm}^2\ \textrm{s}^{-1}$.  Some loss of luminosity (60\% at most) is expected from the anti-pinching~\cite{Zimmermann:1997bh} that takes place between the fully unscattered 70~GeV $e^-$ beam and the 40\% of the electrons in the Compton-scattered beam that escape scattering.   Clues to the additional lost $e^- \gamma$ luminosity can be found in Table~\ref{tab:lumieminussummary}, where four different initial states have total luminosities greater than the geometric luminosity, and two of them -- $e^-e^+$ and $\gamma e^+$ -- contain positrons.  This is not what one expects from $e^-e^-$ collisions, where the  anti-pinch effect should be reducing luminosities with respect to the geometric luminosity. 
 
 The electromagnetic field of the tightly focused 70~GeV electron beam leads to a large amount of coherent pair production by the electrons and photons of the opposing beam.  So many positrons are produced that pinching occurs between the electrons and positrons, which further increases the magnitude of the electromagnetic field, leading to even greater positron production and pinching in a feedback manner.  With symmetric $x$ and $y$ emittances of  120 nm-rad each, fields as high as $4\times 10^{15}$~V/m are produced in the collision.   
 
\begin{figure}
\centering
     \includegraphics[width=0.95\textwidth]{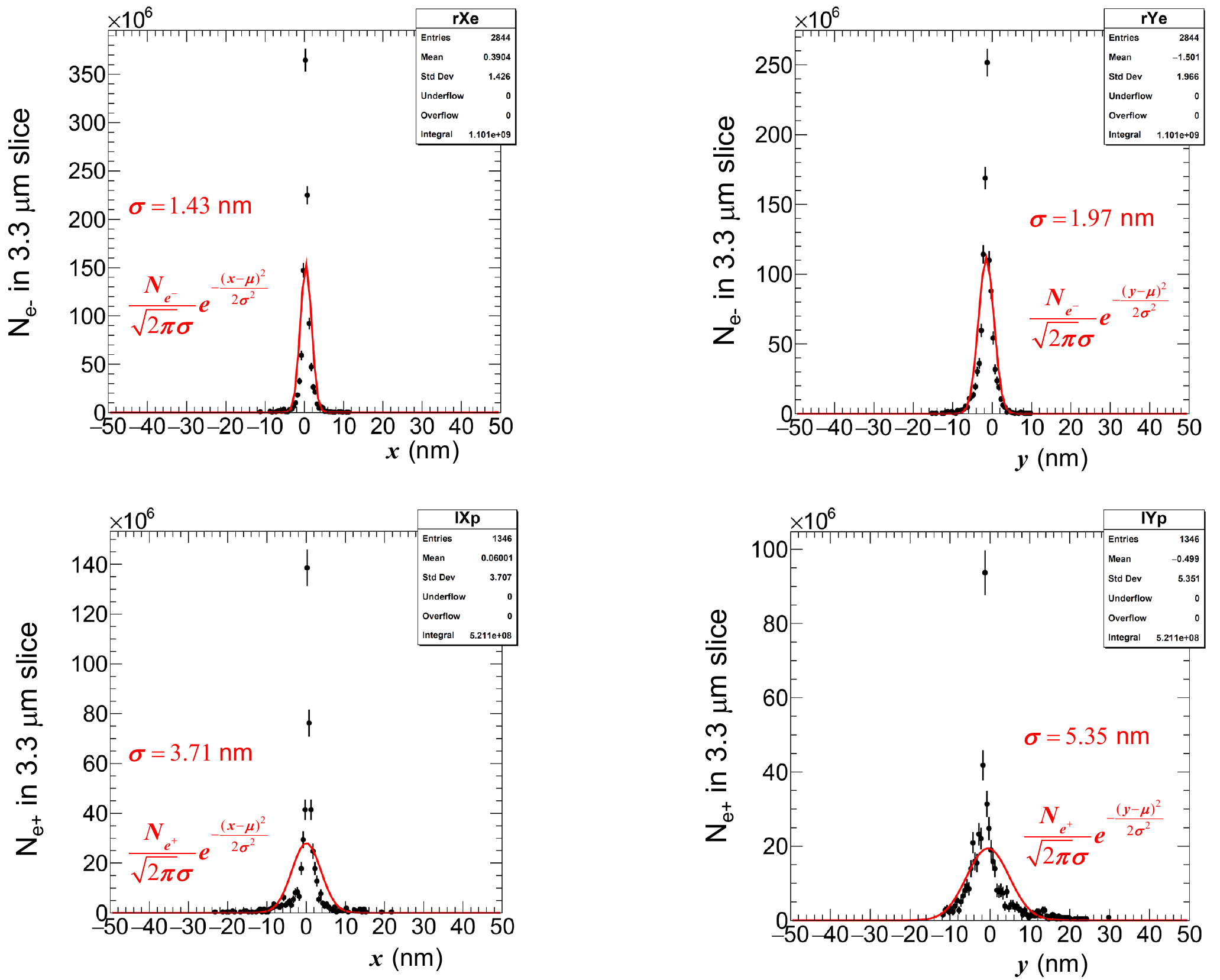}
     \caption{ Snapshots taken about halfway through a $\sqrt{s}=140$~GeV $e^- \gamma$ collision of  particle density distributions for longitudinal slices  of the unscattered 70~GeV $e^-$ beam (top) and positrons in the opposing beam (bottom).  Horizontal (vertical) particle densities are shown on the left (right). Symmetric $x$ and $y$ emittances of  120 nm-rad each were used for the $e^-$ beams. Gaussian fits to the distributions are shown in red.  The $e^-$ beam geometric RMS width for both $x$ and $y$ is 5.1~nm at the IP.}
     \label{fig:posipinch}
 \end{figure}

 Fig.~\ref{fig:posipinch} contains snapshots taken about halfway through a $\sqrt{s}=140$~GeV $e^- \gamma$ collision of  particle density distributions for longitudinal slices  of the unscattered 70~GeV $e^-$ beam and positrons in the opposing beam.  For the $e^-$ distribution, the widths of the Gaussian fits are less than half the nominal width of 5.1~nm, and even narrower non-Gaussian cores are present.  Non-Gaussian cores with widths $\ll 5.1$~nm are also present in the positron distributions.
 
 The solution to large beamstrahlung is to go to asymmetric emittances.  The $e^- \gamma$ baseline invariant emittances of
 $\epsilon_x/\epsilon_y=1200/12$~nm-rad provide a factor 3.8 improvement in event rate over that obtained with the symmetric emittance configuration of 
$\epsilon_x/\epsilon_y=120/120$~nm-rad.

The physics running in $\gamma\gamma$ mode at $\sqrt{s}=125$~GeV and $e^- \gamma$ mode at $\sqrt{s}=140$~GeV does not have to be interleaved. The entire $0.5\times 10^6$ Higgs boson program at $\sqrt{s}=125$~GeV can be performed before moving to 
$\sqrt{s}=140$~GeV.   The 125~GeV program would include energy scans to verify that 
the total Higgs width $\Gamma_{tot}<10$~MeV. This is important to check because the 140~GeV program may not be necessary if the total Higgs width were a few 10's of MeV.

So there would be time to improve the $e^- \gamma$ luminosity. Since asymmetric emittances are used in the baseline configuration, perhaps a damping ring combined with higher currents could help.  Perhaps an IP plasma could be used 
to neutralize the IP~\cite{Zimmermann:1997bh}.  Studies using CAIN indicate that the introduction of an additional 10~GeV $e^-$ beam with suitable timing and location could deflect the Compton-scattered beam just enough to significantly suppress beamstrahlung and coherent pair-production.   

 \newpage

\section{High Brightness LCLS-NC Demonstrator Project}
The required key technologies for the XCC are:
\begin{enumerate}
    \item 1 nC high brightness cryogenic RF gun
    \item 700 mJ/pulse XFEL
    \item 1 keV X-ray optics with 700 mJ/pulse
    \item C$^3$ Acceleration of GeV-class electron beam
\end{enumerate}

Key XCC technology (4) is discussed in the
C$^3$ demonstration document~\cite{Nanni:2022oha}.

Key XCC technologies (1), (2), and (3) can be tested by building a 1 nC high brightness cryogenic RF injector for LCLS-NC~\cite{frisch:2021}.  The construction of such an injector would help demonstrate
key technology (1).  When such an injector is incorporated into LCLS-NC, the upgraded x-ray laser could be used to demonstrate key technologies~(2) and (3).  Furthermore, such a high brightness upgrade to LCLS-NC could open up exciting new research opportunities in photon science.

\subsection{1nC 120 nm-rad cryogenic RF electron gun}
A cryogenic copper RF gun operating at C-band with 0.1 nC/pulse and 45 nm emittance is being developed for an ultra-compact XFEL\cite{Robles:2021ixk}. Scaling arguments indicate that an injector with 1nC/pulse and 120~nm emittance should be possible.  A design study performed several years ago for a cryogenic copper RF gun operating at S-band (TOPGUN) demonstrated that an emittance of 200~nm-rad could be achieved with 1~nC/pulse~\cite{Cahill:2017kwh}\cite{Cahill:2017thesis}.  The development of this gun is a prerequisite for the LCLS-NC high brightness upgrade discussed in the following section. Additional details on high current cryogenic copper RF gun development can be found in the
C$^3$ demonstration document~\cite{Nanni:2022oha}.

\subsection{LCLS-NC performance}
\begin{figure}
\centering
     \includegraphics[width=0.65\textwidth]{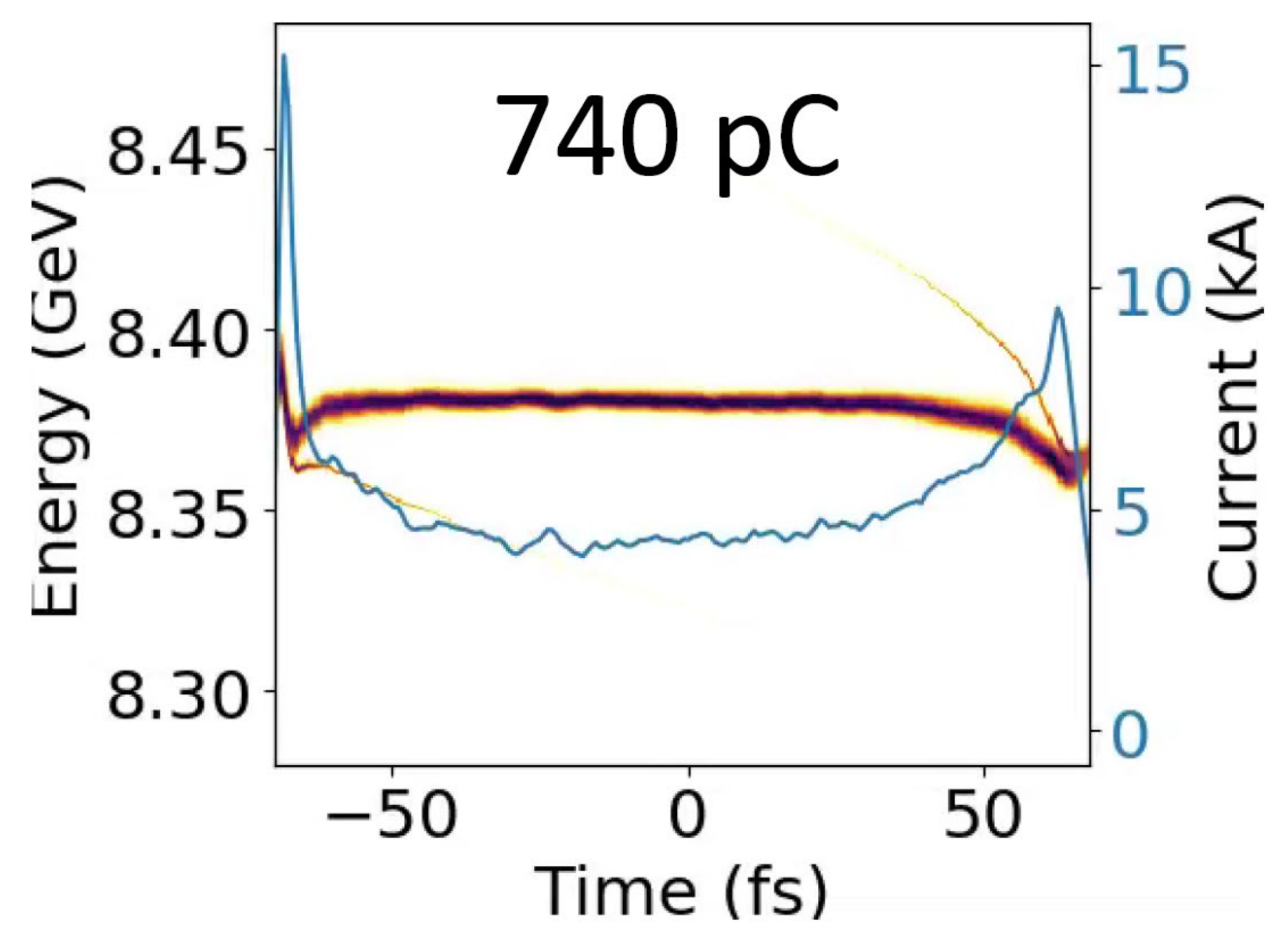}
     \caption{Electron energy and current profile at Linac exit in the high brightness upgrade of LCLS-NC.}
     \label{fig:energycurrentexit}
 \end{figure}

The performance of LCLS-NC assuming an injector with 1~nC charge per pulse,  120 nm-rad emittance and $40\ \mu$m bunch length has been simulated using  ELEGANT\cite{osti_761286} for the Linac and GENESIS for the soft x-ray undulator.  The beam after the injector was created by taking a typical 250~pC beam, scaling the projected emittance from 0.49 to 0.12~$\mu$m in each plane, and increasing the charge per simulation macro particle.  The 1 nC bunch was collimated to 0.74 nC in the first bunch compressor to cut horns.  Linac phases were adjusted to accelerate the beam to 4.5 GeV before the second bunch compressor and compress the beam to 4.5~kA. The compressed beam was accelerated on crest in the first 61 cavities of the third linac to achieve a energy of 8.4~GeV, and the final 117 cavities were set to -80.6 degrees to remove a 15 MeV chirp across the beam. The electron energy and current profile at the exit of the Linac is shown in Fig.~\ref{fig:energycurrentexit}.

Seeded FEL simulations assume a monochrome 50~kW, 1~keV seed approximating a seed from self-seeding to achieve a high brightness x-ray source.  Upon entering the soft x-ray undulator line, the electron beam energy is 8.4~GeV, which is the highest energy beam that can lase at 1~keV in this undulator line (constant period of 3.9~cm and max undulator normalized vector potential of 5.7).  The beam's horizontal and vertical projected emittances of 1.9~$\mu$m and 0.34~$\mu$m, respectively, are dominated by contributions from the horns at the head and tail of the beam, yet the projected core emittance (estimated from the middle 20~fs of beam) remains 0.11~$\mu$m in both planes.  With the standard undulator FODO lattice, the electron beam transverse rms of $11\ \mu$m implies an x-ray waist of $22\ \mu$m, and Rayleigh length of 1~m.  This implies significant diffraction within a 0.8~m gain length (estimated via \cite{Xie2000}), and a shot noise power of 7~kW which is 14\% of the desired seed power of 50~kW which may lead to significant SASE breakthrough. Reducing the FODO quad gradients to 21\% of normal increases the electron beam horizontal rms width to $42\ \mu$m, implying an x-ray Rayleigh range of 5~m, which is significantly longer than the 1.2~m gain length. This also reduces the shot noise to 2.5 kW, or 5\% of the seed power. 

\begin{figure}
\centering
     \includegraphics[width=0.65\textwidth]{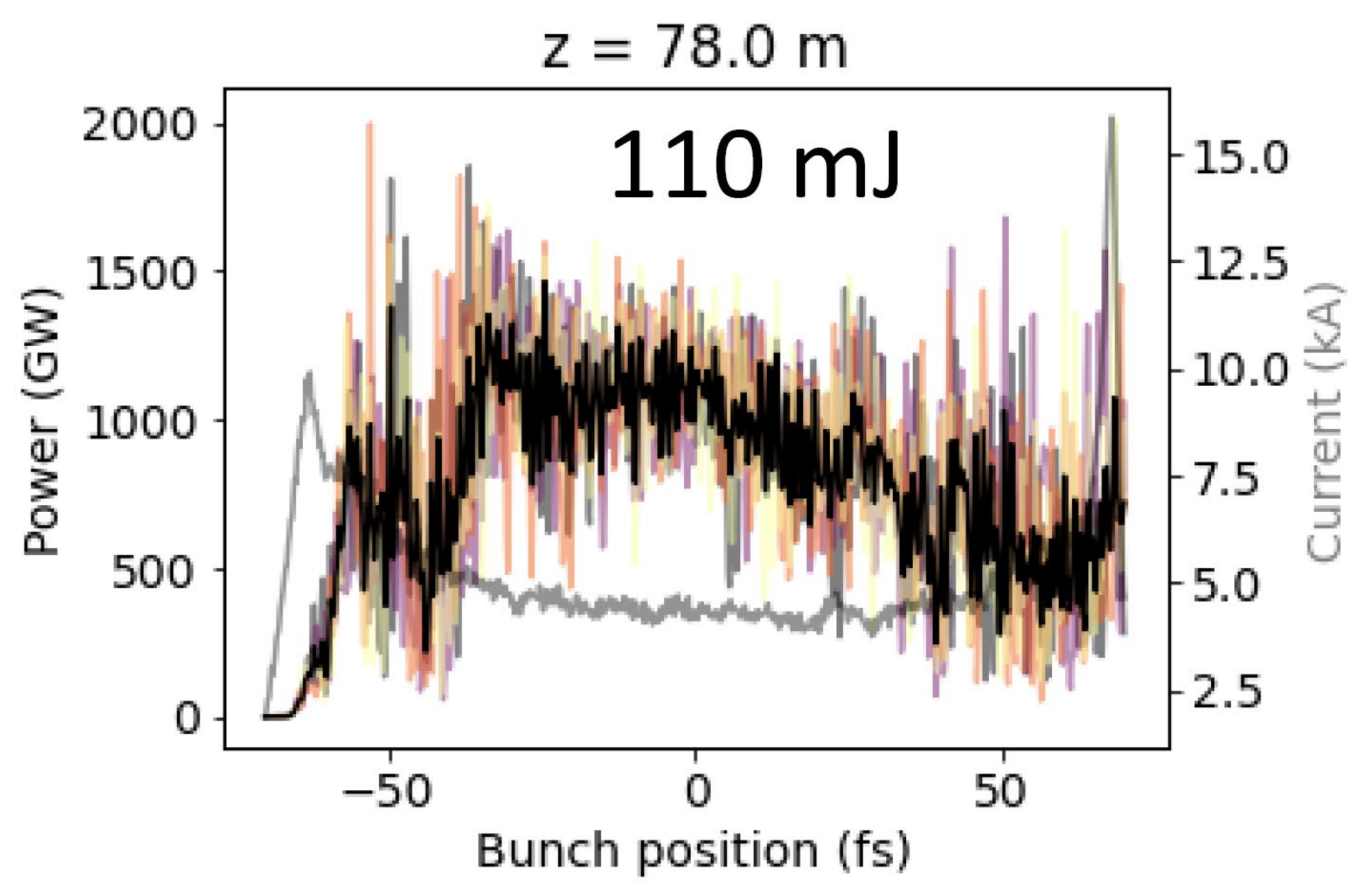}
     \caption{X-ray power profile in the high brightness upgrade of LCLS-NC assuming 21 undulators downstream of the SXRSS chicane.}
     \label{fig:xraypower}
 \end{figure}

\begin{figure}
\centering
     \includegraphics[width=0.65\textwidth]{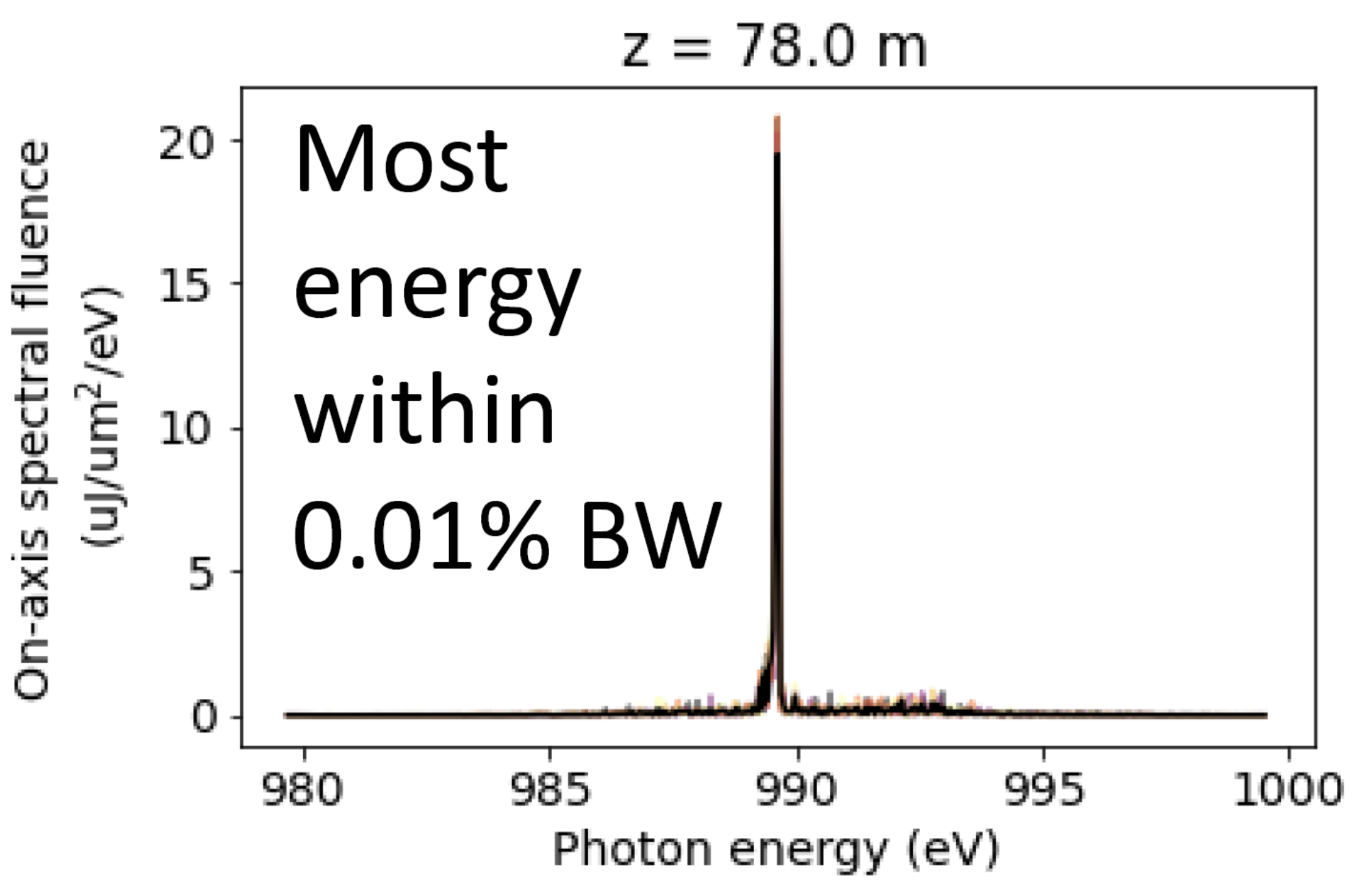}
     \caption{X-ray spectral fluence in the high brightness upgrade of LCLS-NC assuming 21 undulators downstream of the SXRSS chicane.}
     \label{fig:spectralfluence}
 \end{figure}

The seed power for Soft X-ray Self Seeding (SXRSS) is limited to 50~kW to avoid damage to the spectral collimating optics in the SXRSS monochromator. The seeded FEL saturates in about 6 undulators (23.4~m), after which the remaining undulators are quadratically tapered to reduce the undulator field strength by 5.6\% to enhance the peak power to over 1~TW.  After a total of 82~m (21 undulators following the SXRSS chicane) the energy per pulse is 110~mJ with the power profile shown in Fig.~\ref{fig:xraypower}. With the current 12 undulators following the SXRSS chicane, the energy per pulse is 33~mJ. The spectral fluence is shown in Fig.~\ref{fig:spectralfluence} where the bandwidth is less than 0.01\%~FWHM.    

\begin{figure}
\centering
     \includegraphics[width=0.95\textwidth]{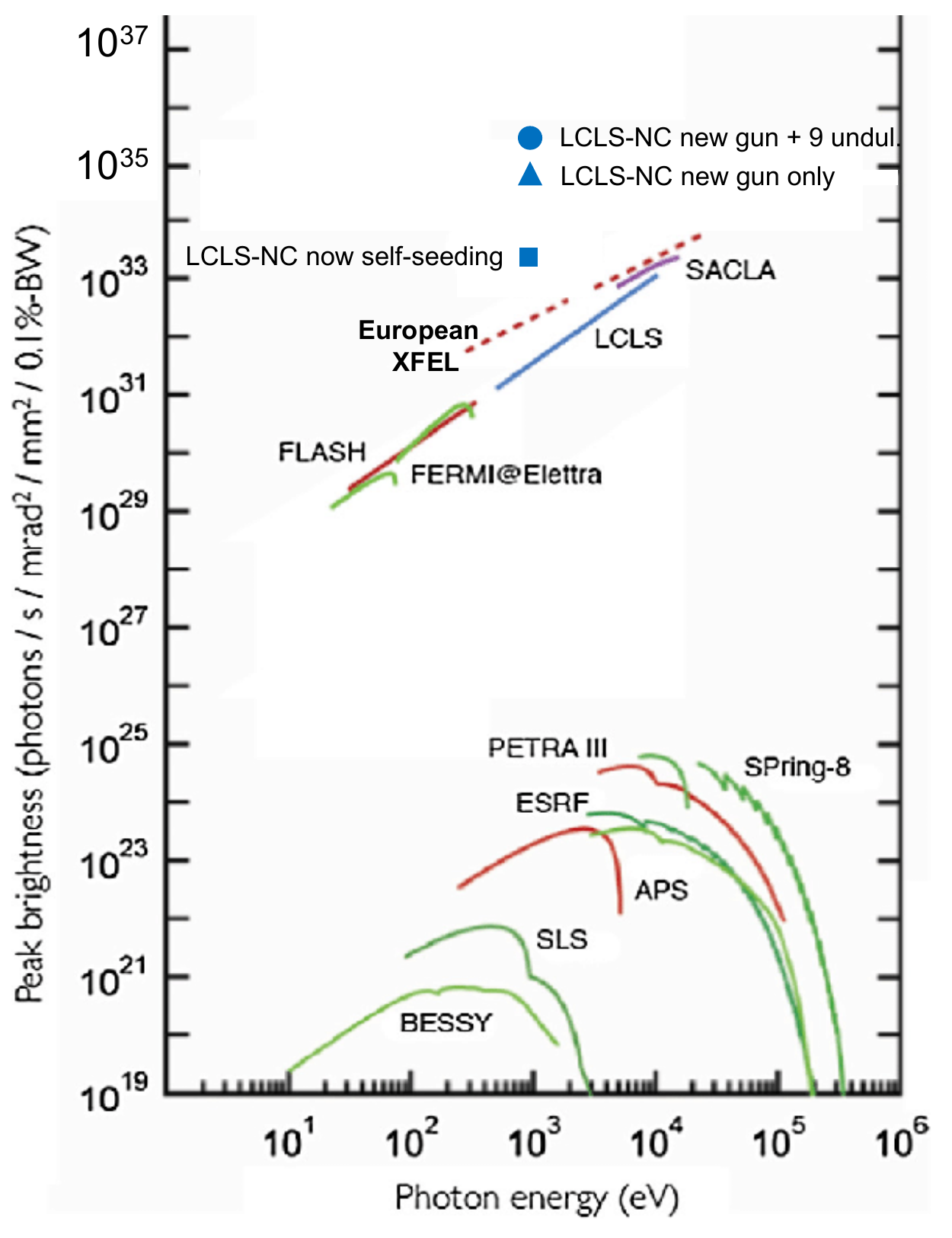}
     \caption{Peak brightness for 1 keV photons of the high brightness upgrade of LCLS-NC in comparison with the peak brightness of other XFEL's. The blue triangle marker indicates the expectation assuming 
     the 1 nC, 120 nm-rad gun cryogenic copper gun only.  The solid blue circle indicates the expectation if in addition nine undulator segments are added to the LCLS soft X-ray undulator line.}
     \label{fig:brightnesscomparison}
 \end{figure}

The 0.01\% bandwidth is much smaller than that required for XCC optical studies, but is important for photon science applications.   Most photon science applications require a much shorter pulse of 10~fs or less.  This can be accomplished with a slotted foil in a dispersive region, with a linear loss in pulse energy versus pulse length.  Furthermore, enhanced SASE could also benefit from low emittance, high current beams to enhance the power of sub-femtosecond soft X-ray pulses.  The peak brightness of the high brightness LCLS-NC upgrade is shown in Fig.~\ref{fig:brightnesscomparison} along with 
the brightness of existing XFEL's.

\newpage

\section{Staging Options and Upgrades}
\subsection{Energy Upgrades}
If the energy of the XCC were upgraded to $\sqrt{s}=280$~GeV, then the XCC could be used to study the Higgs potential through $\gamma\gamma \rightarrow HH$.   An initial study of this prospect has been performed assuming an optical laser~\cite{Kawada:2012uy}. By the time of the XCC energy upgrade the Linac gradient should be 120~Mev/m, so the footprint of XCC at $\sqrt{s}=280$~GeV would be $\sim2.8$~km. The cross section for Higgs pair production in $\gamma\gamma$ collisions at $\sqrt{s}=280$~GeV is about the same as the cross section for $e^+e^-\rightarrow ZHH$ at $\sqrt{s}=500$~GeV.  Without a significant difference in cross section, a detailed study of $\gamma \gamma \rightarrow HH$ is required to determine the relative Higgs self-coupling sensitivity of the XCC and a 500~GeV $e^+e^-$ collider.

Additional energy upgrades could be considered. For example, prolific top quark pair-production occurs for $\sqrt{s}>350$~GeV.

\subsection{Luminosity Upgrade/Staging}
As discussed in Sec~\ref{sec:higgsphysics}, the XCC must upgrade its luminosity by a factor of 3.8 to achieve ILC-like Higgs coupling precision.  However, a large portion of this luminosity upgrade is being used for $e^- \gamma \rightarrow e^- H$ production in $e^- \gamma$ collisions at $\sqrt{s}=140~GeV$. If a significant improvement can be made to the $e^- \gamma$ luminosity through, for example, one of the methods discussed in Sec~\ref{egammachallenge}, then the luminosity for $\gamma\gamma$ collisions at $\sqrt{s}$=125~GeV need only be doubled.

\section{Synergies with other concepts and/or existing facilities}
The XCC would have strong particle physics synergies with LHC.  If  XCC and an $e^+e^-$ collider were both built, there would be clear particle physics synergies between XCC and the $e^+e^-$ collider.  In the nearer term, there are strong synergies between and XCC, C$^3$ and LCLS in the development of the accelerator technology required to realize the XCC or C$^3$-250.

\section{Acknowledgements}

This work was supported in part by the U.S. Department of Energy (DOE) (Contract No. DE-AC02-76SF00515).

%\bibliographystyle{atlasnote}
%\bibliography{bibliography.bib}

%====== Refs ========
\newpage

\printbibliography

% see details of references in the biblatex file attached. Add any references you need there.

\end{document}